\documentclass[12pt,a4paper]{article}
\usepackage{setspace}
\usepackage[pdftex,colorlinks=true]{hyperref}
\usepackage{amssymb,booktabs,subfig,xcolor,microtype,paralist,fullpage, makecell, longtable, bm, orcidlink, dsfont}
\usepackage[top=1in, bottom=1in, left=1in, right=1in]{geometry}
\usepackage{natbib}
\usepackage{url} 
\definecolor{darkblue}{rgb}{0,0,.6}
\hypersetup{citecolor=darkblue,linkcolor=darkblue,urlcolor=darkblue}
\usepackage{amsmath}
\usepackage{amsfonts}
\usepackage{epsfig}
\usepackage{graphics}
\usepackage{palatino,eulervm}
\usepackage{graphicx}
\usepackage{lscape}
\usepackage{rotating}
\usepackage[linewidth=1pt]{mdframed}
\allowdisplaybreaks[4]
\newcommand{\blind}{0}

\newcommand{\X}{\mathcal{X}}

\newcommand{\Rlogo}{\protect\includegraphics[height=1.8ex,keepaspectratio]{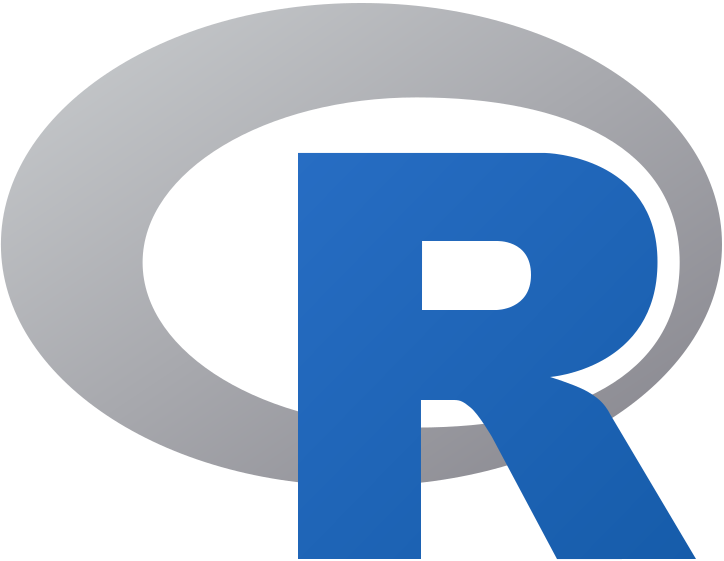}}
\newsavebox\CBox

\providecommand{\U}[1]{\protect\rule{.1in}{.1in}}
\renewcommand{\baselinestretch}{1.2}
\graphicspath{{plots/}}

\usepackage{amsthm,thmtools}
\usepackage{mathrsfs}

\makeatletter
\def\th@newremark{\th@remark\thm@headfont{\bfseries}}
\makeatletter
\theoremstyle{newremark}

\declaretheoremstyle[
  spaceabove=6pt, spacebelow=6pt,
  headfont=\bfseries,
  notefont=\mdseries, notebraces={(}{)},
bodyfont=\normalfont,
  postheadspace=0.5em,
]{mystyle}

\begin{document}

\def\spacingset#1{\renewcommand{\baselinestretch}
{#1}\small\normalsize} \spacingset{1}

\if0\blind
{
\title{\bf Is the age pension in Australia sustainable and fair? Evidence from forecasting the old-age dependency ratio using the Hamilton-Perry model}

\author
{
Sizhe Chen\orcidlink{0009-0002-5062-2951} \qquad Han Lin Shang\orcidlink{0000-0003-1769-6430} \\
Department of Actuarial Studies and Business Analytics \\
Macquarie University \\
\\
Yang Yang \orcidlink{0000-0002-8323-1490}  \\
School of Information and Physical Sciences \\
The University of Newcastle \\
}
\date{}
\maketitle
} \fi

\if1\blind
{
\title{\bf Is the age pension in Australia sustainable and fair? Evidence from forecasting the old-age dependency ratio using the Hamilton-Perry model}
\maketitle
} \fi

\bigskip
\begin{abstract}
The age pension aims to assist eligible elderly Australians who meet specific age and residency criteria in maintaining basic living standards. In designing efficient pension systems, government policymakers seek to satisfy the expectations of the overall aging population in Australia. However, the population's unique demographic characteristics at the state and territory level are often overlooked due to the lack of available data. We use the Hamilton-Perry model, which requires minimum input, to model and forecast the evolution of age-specific populations at the state level. We also integrate the obtained sub-national demographic information to determine sustainable pension ages up to 2051. We also investigate pension welfare distribution in all states and territories to identify disadvantaged residents under the current pension system. Using the sub-national mortality data for Australia from 1971 to 2021 obtained from \cite{AHMD23}, we implement the Hamilton-Perry model with the help of functional time series forecasting techniques. With forecasts of age-specific population sizes for each state and territory, we compute the old age dependency ratio to determine the nationwide sustainable pension age.

\vspace{.1in}

\end{abstract}

\spacingset{1.4}

\newpage

\section{Introduction}\label{sec:1}

Population aging presents long-term economic and fiscal challenges to the sustainability of the social welfare system in Australia. The Australian \textit{Intergenerational Report} \citep{Crescent23} indicates that the number of people aged 65 and over will at least double over the next 40 years. The notable increase in life expectancy seen in developed countries like Australia in recent decades can be attributed to advancements in healthcare, better living standards, and more effective strategies in promoting health and preventing diseases. \cite{GRS+14} analyzed the data and described probabilistic population projections for the entire world, indicating the ratio of working-age people to older people is almost certain to decline substantially in all countries. 

In Australia, citizens born after 2010 enjoy a life expectancy of over 80 years, around a 40\% increase since the start of the $20^{\text{th}}$ century \citep{AIHW22}. This extended longevity brings additional risks to the Australian superannuation system as more citizens may have insufficient savings to support them through their retirement without recourse to the age pension \citep{Sup22}. 

Government stakeholders seek to design efficient social payment systems, which requires accurately forecasting longevity risk. However, there is a critical problem: a need for more reliable age-specific demographic data reflecting the influence of remoteness on residents, especially the indigenous population, in rural and remote areas in Australia \citep{AIHW03}. For instance, there have been concerns about the accuracy of mortality data for rural and remote jurisdictions where a proportion of the registered deaths of Aboriginal and Torres Strait Islander peoples are not correctly identified as such \citep{ABS22}. Moreover, data availability makes modeling internal migration between states and territories difficult. People may move from rural to urban areas or between cities based on job prospects, industry growth, economic development, and environmental factors. In addition, frequent changes in immigration policies significantly impact the number of international immigrants in each state and territory. As a result, poor population data quality at the sub-national level poses great challenges for designing an effective age pension system.

We propose modeling and projecting state and territory populations using the Hamilton-Perry (HP) model \citep{HP62} and use the forecasts to determine the sustainable pension age. The HP model overcomes the restriction of data availability at the sub-national level since it requires significantly fewer inputs and computational time to implement than the traditional cohort-component method \citep[see, e.g.,][]{Shang2016} or multiregional population projection \citep[see, e.g.,][]{Rogers75}. Specifically, the standard cohort-component approach for population projection requires high-quality data about mortality, fertility, and immigration of a population and assumptions. In contrast, the HP model only requires population size by age and sex from the two most recent censuses \citep[][Chapter 7]{STS01}. This has made the HP model popular over the last decade in small area population forecasts, with notable applications including an initial population projection for Clark County, Nevada \citep{SSS10} and a subsequent nationwide census tract evaluation across all sub-county areas in the United States \citep{BST21}, to name only a few. Most recently, a modified HP model was considered to forecast populations for small areas of Australia at the statistical area Level 2 scale \citep[defined as in][]{ABS2011}. This modified HP model with smoothed age profiles produces the most accurate forecasts \citep{WG22}. 

This paper contributes a new variant of the HP model by extending the idea of a smoothing cohort change ratio (CCR, \citeauthor{WG22}, \citeyear{WG22}) in combination with functional time series forecasting techniques \citep[see][for an introduction to functional data analysis]{RS05, FV06}. Functional time series methods have been widely used in demographic data modeling and forecasting \citep[see, e.g.,][]{CBD+11, DPR11, CCO17}. Using this framework, the CCR of sex $s$ increasing from age $x$ to $x+1$ within the time interval $[t-1, t]$, $\text{CCR}^s_{x\rightarrow x+1}(t-1\rightarrow t)$, can be regarded as a stochastic process defined on a compact set $x \in \mathcal{I}$ and indexed by a time label $t \in \mathbf{Z}$. This functional time series setting is best known for retaining the correlation of demographic observations at different ages, which enables extracting the temporal dependence of the underlying stochastic process for forecasting purposes. These advantages allow functional time series methods to produce more accurate forecasts than the commonly used parametric models \citep[see, e.g.,][]{HU07}. With minimal data input, we combine functional time series forecasting techniques with the HP model to produce accurate sub-national population forecasts. By aggregating the sub-national population forecasts to the national level, we determine a sustainable pension age based on the current old-age dependency ratio.

As of 31 March 2023, around 2.8 million Australians aged 65 and over received income support payments, with the overwhelming majority (92\%, or 2.6 million individuals) receiving the age pension \citep{AIHW23}. The impacts of recent increases in Australian pension entitlement ages on the health and living conditions of residents are not completely clear \citep[see, e.g.,][]{Atalay14}. Differences in socioeconomic status contribute to inequalities in life expectancies between metropolitan and rural residents \citep{Clarke11, Jacobs18}. Under Australia's current noncontributory pension scheme, eligibility does not require prior employment, and the beneﬁt levels are not determined by prior earnings or residence locations. The uniform pension rate may not provide enough support to residents in regional and rural areas where access to medical services is more difficult and expensive than in metropolitan cities. Using sub-national mortality forecasts produced by the HP model, we compute life expectancies for residents in all states and territories. With the forecast pension rate for singles and couples, we find out that residents in the Northern Territory and Tasmania are expected to receive the least age pension income throughout their entire retirement. In contrast, retirees living in New South Wales and Victoria are expected to enjoy the highest expected age pension income due to longer life expectancies. The large gaps in retirees' lifetime age pension income indicate disadvantages in the current welfare distribution program for pensioners. To reduce inequalities in income distribution, the Federal Government recommends offering more help to local governments in the Northern Territory and Tasmania to provide more health services accessible to most pensioners in regional and rural areas.

The remainder of the paper is organized as follows. Section~\ref{sec:2} describes the age- and sex-specific population data at Australia's national and sub-national levels. Section~\ref{sec:3} provides details of the population forecasting method. The empirical data analysis results are presented in Section~\ref{sec:4}. Section~\ref{sec:5} proposes an approach to determine the optimal pension age based on the forecasted population and its old-age dependency ratio. With the help of the HP model, we study the changes in population at the sub-national level. From the historical observed data, we generate out-of-sample population forecasts from 2022 to 2051 and determine a sustainable pension age for Australia. With the sub-national population projections and forecasts of pension payout rates, we compare lifetime age pension income for residents in Australia's six states and two territories in Section~\ref{sec:6}. Finally, Section~\ref{sec:7} concludes the paper and provides some directions for future research. 
 
\section{Population data in Australia}\label{sec:2}

\subsection{National age- and sex-specific data}\label{sec:2.1}

The Australian age- and sex-specific population data from 1921 to 2021 are publicly available from the \cite{HMD23} for each state. The data consists of population counts for every single year of age, ranging from 0 to 99, with the last group being $100+$. By studying the changes in population concerning age and year, we can observe that there has been an upward trend in population over time. 

\begin{figure}[!htb]
\centering
\subfloat[Female population in Australia]
{\includegraphics[width=8.5cm]{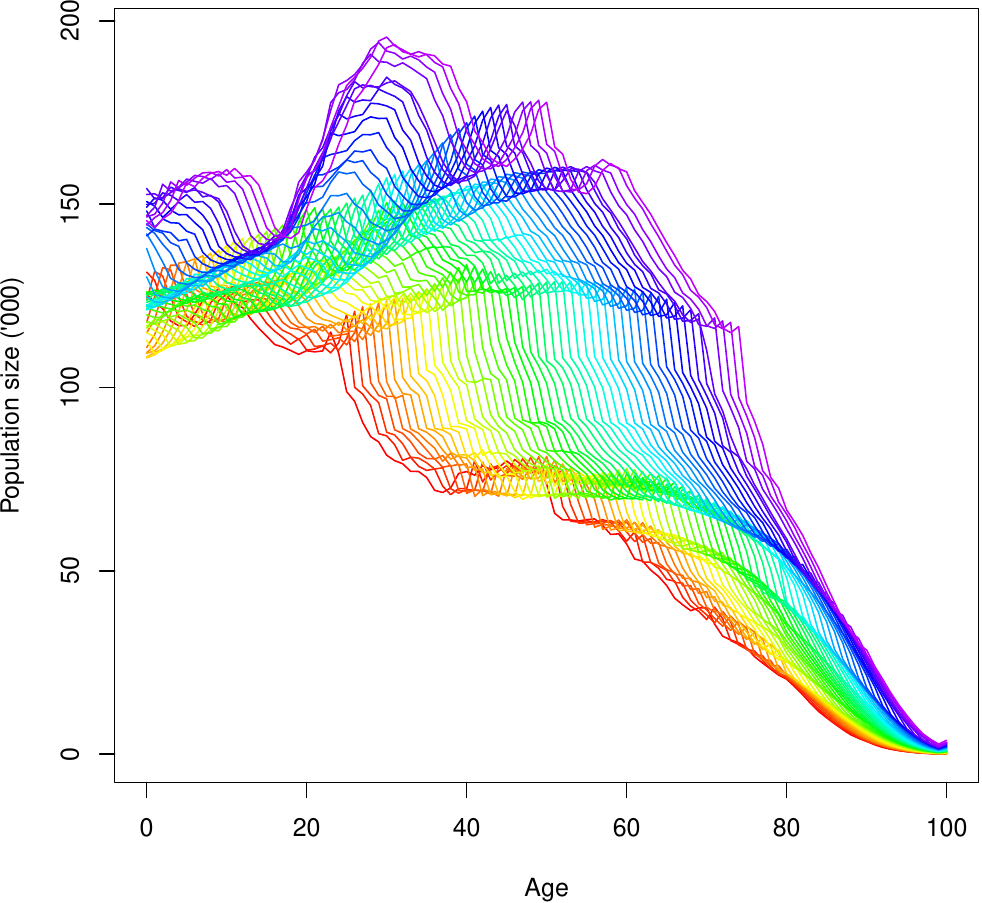}}
\quad
\subfloat[Male population in Australia]
{\includegraphics[width=8.5cm]{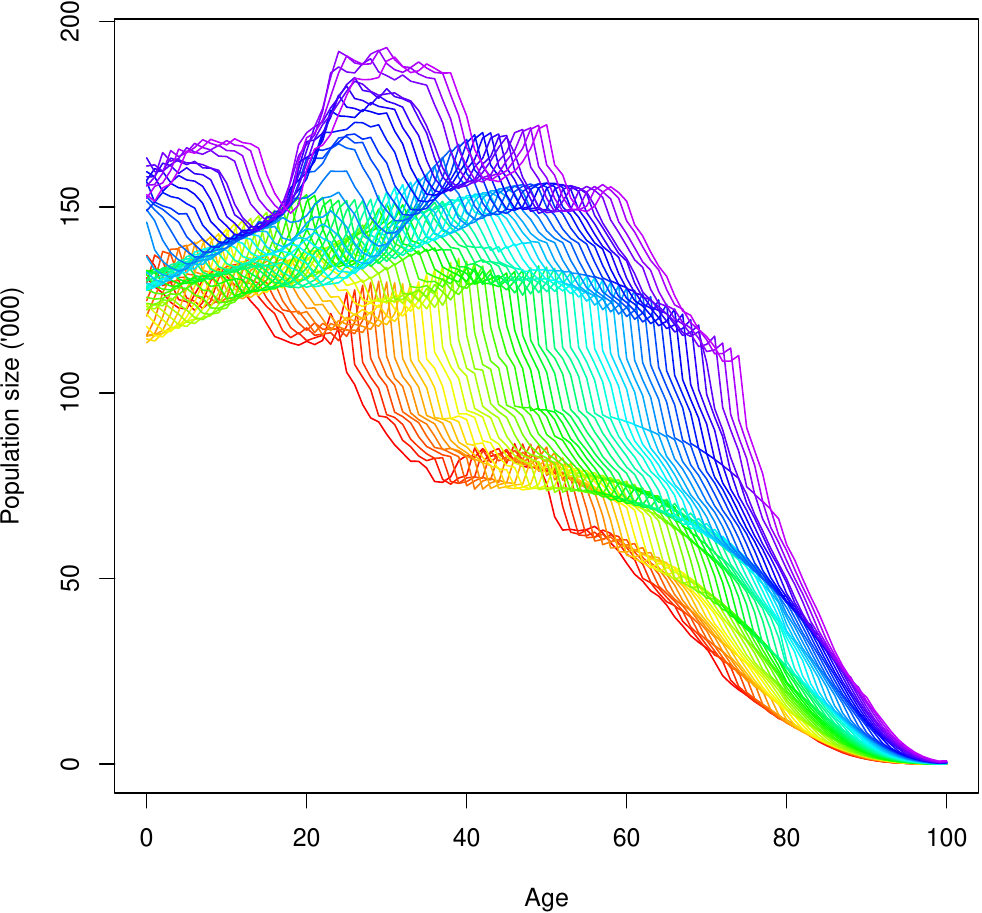}}
\caption{\small{Rainbow plots of Australian female and male population sizes from 1971 to 2021 between ages 0 and 99 in single years of age, with the last age group being 100+}}\label{fig:1}
\end{figure}

To better represent this evolution, we have combined the population counts from each state and used rainbow plots in Figure~\ref{fig:1} to show the total population sizes, with the population from the distant past shown in red and the most recent years depicted in purple. In recent years, Australia has seen relatively stable birth and death rates. However, international migrants between the ages of 20 and 50 have played a significant role in population growth and demographic changes.

\subsection{Sub-national age- and sex-specific data}\label{sec:2.2}

While national-level data provides an overall picture of a country's demographic trends, economic indicators, and other key statistics, sub-national details are crucial for understanding regional variations and making more targeted policy decisions. Sub-national data allow for granular analysis of regional disparities, urban-rural differences, resource allocation, and infrastructure planning.

\begin{figure}[!htb]
\centering
\includegraphics[width=8.8cm]{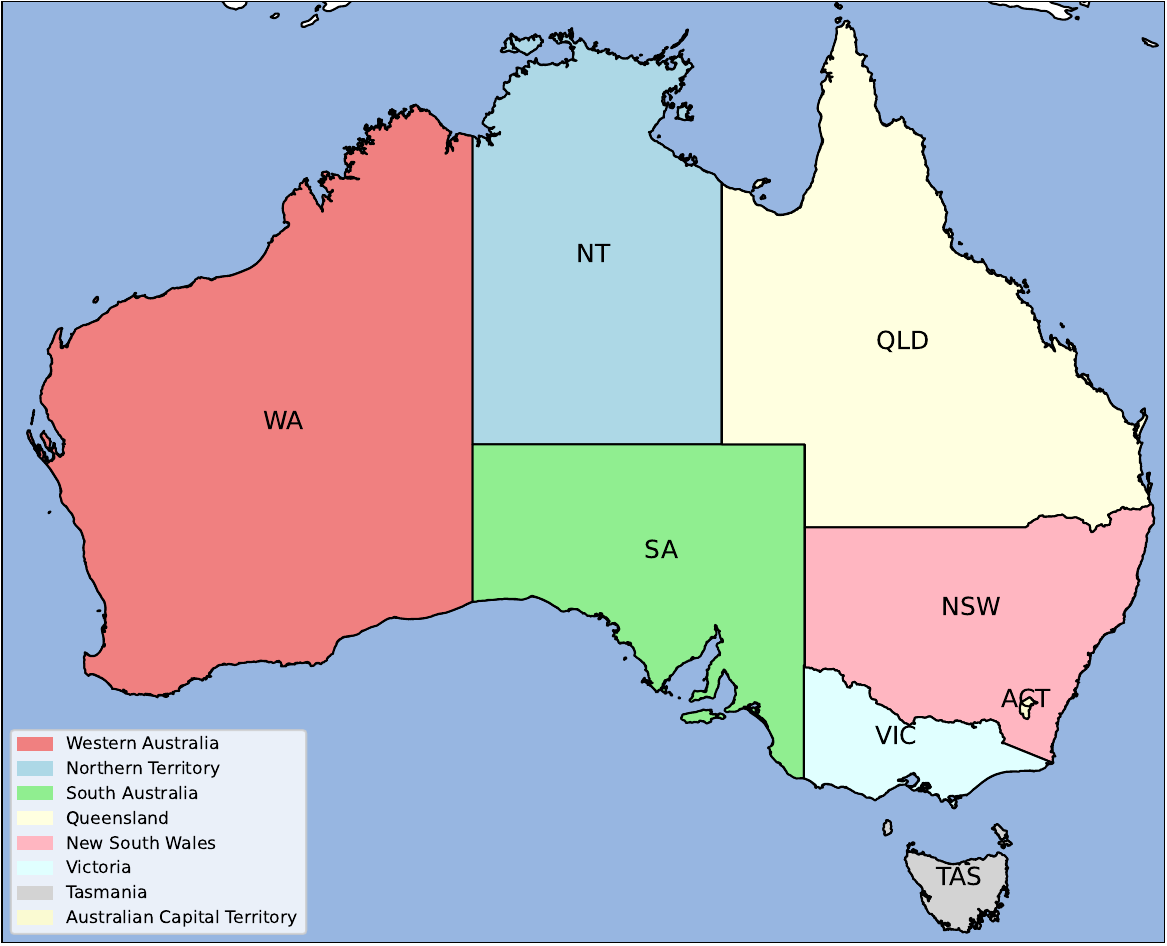}
\caption{A map of six Australian states and two territories.}\label{fig:1.5}
\end{figure}

The Australian age-specific sub-national data from 1971 to 2021 are available from \cite{AHMD23}. We consider the eight first-level administrative divisions in Australia, namely the Australian Capital Territory (ACT), New South Wales (NSW), Queensland (QLD), Western Australia (WA), the Northern Territory (NT), Tasmania (TAS), Victoria (VIC) and South Australia (SA). These regions are displayed in Figure~\ref{fig:1.5}. Population estimates are provided by single years of age up to 109, with an open age interval for 110+. As the population sizes at comparably higher ages are almost all 0, we consider the population counts for single years of age from 0 to 99, with the last age group being 100+. This grouping of ages aligns with the national-level data.

\section{Population forecasting}\label{sec:3}

\subsection{Hamilton-Perry model}\label{sec:3.1}

A recent study by \cite{wilson2021brief} outlined the standard process for creating a national population forecast using a cohort-component model. This model requires high-quality data on fertility, mortality, and internal and international migration. The quality of these datasets needs to be carefully examined to ensure accurate population forecasts. The cohort-component model is advantageous as it effectively represents underlying population processes and provides age-disaggregated projections for various purposes such as government planning, policy-making, budgeting, and service delivery. However, it can also be highly inaccurate if poor assumptions are made about fertility, mortality, and migration. Additionally, the cohort-component model requires significant input data, and assumptions about future fertility, mortality, and migration must be prepared separately. 

\begin{figure}[!htb]
\begin{center}
\begin{tikzpicture}
\filldraw[lightgray, thick] (2,4) -- (0,2) --  (0,0) -- (2,2) -- (2,4);
\node at (0,-0.2){$t-1$};
\node at (2,-0.2){$t$};
\node at (-0.4,1){$P^s_{x}$};
\node at (2.6,3){$P^s_{x+1}$};
\node at (-1,1){$x$};
\node at (-1,3){$x+1$};
\node at (-3, 2){Age group};
\node at (1, -0.6){Year};
\draw[ultra thick] (0,0) -- (0,2);
\draw[ultra thick] (2,2) -- (2,4);
\draw [-latex] (0.5,1.5) -- (1.5,2.5);
\end{tikzpicture}
\caption{\small{Lexis diagram representation of the CCR in the HP model.}}\label{fig:1.6}
\end{center}
\end{figure}
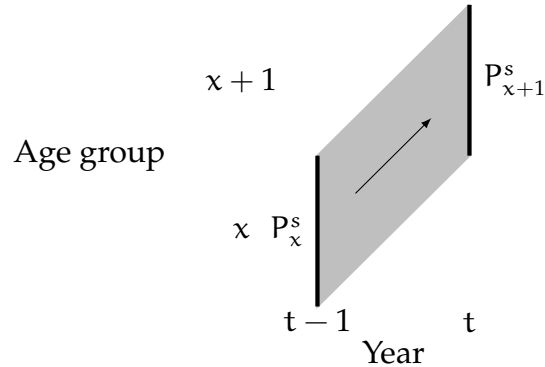

Compared to the cohort-component model, \citeauthor{HP62}'s \citeyearpar{HP62} method overcomes these drawbacks as it requires much smaller data requirements for projection. Instead of modeling mortality, fertility, and migration, the HP model requires data from the two most recent censuses \citep[][pp. 153--158]{STS01}. The ratio between two population censuses is known as the CCR. It measures the combined net effect of mortality and migration experienced by the cohort over time. It is estimated from population sizes by age and sex for the jump-off year and (commonly) one year earlier. For example, the CCR for the cohort of sex $s$ at age interval $x$ at time $t-1$ and age $x+1$ at time $t$ is defined as
\begin{equation*}
\text{CCR}^s_{x\rightarrow x+1}(t-1\rightarrow t) = \frac{P^s_{x+1}(t)}{P^s_{x}(t-1)}, \qquad x\geq 1.
\end{equation*}
The infant age group, that is $0-1$, is forecast differently. Because the infant age group depends on fertility, we consider the child/women ratio (CWR), the ratio of the population aged $0-1$ to the women of childbearing ages from 15 to 49. The population aged $0-1$ is forecast by multiplying the CWR by the women aged 15 to 49. The infant population is then multiplied by the birth sex ratio to yield the number of baby boys or girls. The forecast can be expressed as
\begin{equation}
P_{0}^s(t+1) = P_{\text{F}, 15-49}(t+1)\times \text{CWR}(t+1)\times \text{birth sex ratio}^s(t+1),\label{eq:sex_ratio}
\end{equation}
where the birth sex ratio between infant boy and girl is $1.057$:$1$ in 2021 \citep{ABSbirth}. In Figure~\ref{fig:1.6}, we present a lexis diagram representation of the CCR in the HP model.

The HP model is commonly used for predicting the population of small areas \citep[e.g.,][]{WGA+22}. With the sub-national population forecasts for eight regions, we study the dynamic changes in regional population counts. By aggregating the population forecasts at the national level, we provide Australian Government policymakers with a future sustainable retirement age.

\subsection{Functional time series forecasting}\label{sec:3.2}

Functional data analysis has received increasing amounts of attention in demographic forecasting \citep[see, e.g.,][]{CBD+11, DPR11, CCO17} since its first application to demographic modeling and forecasting by \cite{HU07}. While \cite{HS09} and \cite{HBY13} apply functional time series analysis to age-specific mortality or fertility rates, \cite{HB08} extended this method to the modeling and forecasting of age- and sex-specific population sizes. 

We treat a time series of age-specific populations as a functional time series. Define $\X(u)$ as a stochastic process for age-specific populations, where $u$ represents the continuous age variable within the compact function support range $\mathcal{I}$. Let $\X_t(u)$ denote a collection of realisations of $\X(u)$ over time such that each observation $\X_t$ is an element of the Hilbert space and satisfies $\int_{\mathcal{I}}\X_t^2(u)du<\infty$. In practice, the population observations are reported annually over $u\in [0,100+]$ at $t=1, 2, \dots, n$, resulting in a set of $n$ curves, namely $\{\X_1(u), \X_2(u), \dots , \X_n(u)\}$. Let $\mu$ be the mean function of $\X$, the covariance function of $\X$ is defined as
\begin{align*}
K(u,v) &= \text{Cov}[\X(u), \X(v)] \\
	   &=\text{E}\left\{[\X(u) - \mu(u)][\X(v) - \mu(v)]\right\}.
\end{align*}
The $K(u,v)$ is an analogue of the variance-covariance matrix commonly used in multivariate data analysis. The variation in population functions over the entire domain of the age variable $u$ is captured by the variance function \citep[][Chapter 8]{RS05}.


Via Mercer's lemma, there exists an orthonormal sequence $(\phi_k)$ of continuous function in $\mathcal{L}^2(\mathcal{I})$ and a non-increasing sequence $(\lambda_k)$ of positive numbers, such that
\begin{equation*}
K(u,v) = \sum_{k=1}^{\infty}\lambda_k \phi_k(u)\phi_k(v), \qquad u,v \in \mathcal{I}.
\end{equation*}
Via Karhunen-Lo\`{e}ve expansion, $\X$ can be expressed as
\begin{equation}
\X(u) = \mu(u) + \sum^{\infty}_{k=1}\beta_k\phi_k(u),\label{eq:1}
\end{equation}
where $\beta_k = \langle \X(u) - \mu(u), \phi_k(u)\rangle$ is an uncorrelated random variable with zero mean and unit variance. To reduce dimensionality, we truncate the number of components to the first $K$ term:
\begin{equation*}
\X_t(u) = \overline{\X}(u) + \sum^K_{k=1}\widehat{\beta}_{t,k}\widehat{\phi}_k(u)+e_t(u),
\end{equation*}
where $\overline{\X}(u)=\frac{1}{n}\sum_{t=1}^n\X_t(u)$ denotes the sample mean function, $K$ denotes the retained number of principal components, $\widehat{\beta}_{t,k}$ denotes the $k$\textsuperscript{th} set of the estimated principal component scores, $\widehat{\phi}_k(u)$ denotes the $k$\textsuperscript{th} set of the estimated functional principal component, and $e_t(u)$ denotes the model residual at time $t$.

There are many ways of determining the optimal number of $K$, such as the bootstrap approach of \cite{HV06} and \cite{BYZ10}, the description length approach of \cite{PS13}, pseudo-AIC \citep{Shibata81}, scree plot \citep{Cattell66}, the eigenvalue ratio criterion \citep{LRS20}, and eigenvector variability plot \citep{TCC09}. The functional time-series forecasting method is robust in overestimating $K$, but underestimating $K$ can result in poor performance. The forecasting method is robust in overestimating $K$, but underestimating $K$ can result in inferior accuracy. As in \cite{HBY13}, we select $K=6$.

Conditioning on the estimated mean function and estimated functional principal components $\widehat{\bm{\Phi}} = [\widehat{\phi}_1(u),\widehat{\phi}_2(u),\dots,\widehat{\phi}_K(u)]$, the $h$-step-ahead point forecast can be obtained 
\begin{align*}
\widehat{\X}_{n+h|n}(u) &= \text{E}[\X_{n+h}(u)|\overline{\X}(u),\widehat{\bm{\Phi}}] \\
                        &= \overline{\X}(u) + \sum^K_{k=1}\widehat{\beta}_{n+h|n,k}\widehat{\phi}_k(u),
\end{align*}
where $\widehat{\beta}_{n+h|n,k}$ denotes the $h$-step-ahead forecast of the principal component scores obtained from a univariate time series forecasting method, such as the autoregressive integrated moving average (ARIMA) model. In our application, the described process forecasts the mortality function $\X_t(u)$ for the entire age range $u \in [0, 100+]$. Consequently, the interdependence among different ages is preserved during the forecasting.

\subsection{Construction of prediction intervals via nonparametric bootstrapping}\label{sec:3.3}

Prediction intervals are valuable for evaluating the probabilistic uncertainty associated with point forecasts. Forecast uncertainty stems from systematic deviations (e.g., due to parameter or model uncertainty) and random fluctuations (e.g., due to model error term). As was emphasized by \cite{Chatfield93, Chatfield00}, it is essential to provide interval forecasts as well as point forecasts to:
\begin{enumerate}
\item[(1)] assess future uncertainty levels;
\item[(2)] enable different strategies to be planned for a range of possible outcomes indicated by the interval forecasts;
\item[(3)] compare forecasts from different methods more thoroughly; and
\item[(4)] explore different scenarios based on various assumptions.
\end{enumerate}

There are two sources of uncertainty: truncation errors in the functional principal component decomposition and forecast errors in the forecast principal component scores. Since principal component scores are regarded as surrogates of the original functional time series, these principal component scores capture the temporal dependence structure inherited in the original functional time series \citep[see also][]{Paparoditis18, Shang18, PS23}. By bootstrapping the forecast principal component scores, we can generate a set of bootstrap samples $\X_{n+h}^*(u)$, conditional on the mean function and functional principal components estimated from the observed data $\bm{\X}(u) = [\X_1(u), \X_2(u),\dots, \X_n(u)]$.

Using a univariate time series model, we can obtain multi-step-ahead forecasts for the principal component scores, $\{\widehat{\beta}_{1,k},\dots,\widehat{\beta}_{n,k}\}$ for $k=1,\dots,K$. Let the $h$-step-ahead forecast errors be given by $\widehat{\pi}_{t,h,k}=\widehat{\beta}_{t,k}-\widehat{\beta}_{t|t-h,k}$ for $t=h+1,\dots,n$. These errors can be sampled with replacement to given bootstrap samples of $\beta_{n+h,k}$:
\begin{equation*}
\widehat{\beta}_{n+h|n,k}^{(b)}=\widehat{\beta}_{n+h|n,k}+\widehat{\pi}^{(b)}_{*,h,k}, \qquad b=1,\dots,B,
\end{equation*}
where $B=1,000$ symbolizes the number of bootstrap replications and $\widehat{\pi}^{(b)}_{*,h,k}$ are sampled with replacement from $\{\widehat{\pi}_{1,h,k},\dots,\widehat{\pi}_{n,h,k}\}$.

Suppose the functional principal component decomposition approximates the data $\bm{\X}(u)$ relatively well, the model residual $e_t(u) = \X_t(u) - \overline{\X}(u) - \sum^K_{k=1}\widehat{\beta}_{t,k}\widehat{\phi}_k(u)$ should contribute nothing but random noise.

Adding the two sources of variability, we obtain $B$ variants for $\X_{n+h}(u)$:
\begin{equation*}
\X_{n+h|n}^{(b)}(u) = \overline{\X}(u) + \sum^K_{k=1}\widehat{\beta}_{n+h|n,k}^{(b)}\widehat{\phi}_k(u)+e^{(b)}(u),
\end{equation*}
where $\widehat{\beta}_{n+h|n,k}^{(b)}$ denotes the bootstrap forecast of the principal component scores. With the bootstrapped $\X_{n+h|n}^{(b)}(u)$, the pointwise prediction intervals are obtained by taking $\alpha/2$ and $(1-\alpha/2)$ quantiles at the $100(1-\alpha)\%$ nominal coverage probability and $\alpha$ denotes a level of significance, such as $\alpha=0.2$ or 0.05.

\subsection{Multivariate functional time series method}\label{sec:3.4}

Inspired by joint modeling, a multivariate functional time series method is proposed to model and forecast multiple series simultaneously \citep{SK22}. The multivariate functional time series are stacked in a vector with $\bm{\X}_t(u) = [\X_t^{\text{F}}(u), \X_t^{\text{M}}(u)]$. For $u, v\in \mathcal{I}$, the theoretical cross-covariance function can be defined with elements
\begin{align*}
c_{\text{F, M}}(u,v) :&= \text{Cov}[\X^{\text{F}}(u), \X^{\text{M}}(u)] \\
&=\text{E}\left\{[\X^{\text{F}}(u) - \mu^{\text{F}}(u)][\X^{\text{M}}(u) - \mu^{\text{M}}(u)]\right\}
\end{align*}
where $\mu^{\text{F}}(u)$ and $\mu^{\text{M}}(u)$ denote the mean functions for the female and male series, respectively. Via Mercer's lemma, there exists an orthonormal sequence $(\phi_k)$ and a non-increasing sequence $\lambda_k$ of positive numbers, such that
\begin{equation*}
c_{\text{F, M}}(u, v) = \sum^{\infty}_{k=1}\lambda_k \phi_k^{\text{F}}(u)\phi_k^{\text{M}}(v).
\end{equation*}
By Karhunen-Lo\`{e}ve expansion, a realization of the stochastic process $\bm{\X}(u)$ can be expressed as
\begin{equation*}
\X_t^{\text{F}}(u) = \overline{\X}^{\text{F}}(u) + \sum^K_{k=1}\widehat{\beta}_{t,k}^{\text{F}}\widehat{\phi}_k^{\text{F}}(u),
\end{equation*}
where $K$ denotes the number of retained functional principal components which is the same across multiple series, $\widehat{\phi}_k^{\text{F}}(u)$ represents the $k$\textsuperscript{th} estimated functional principal component corresponding to the female population, and $\widehat{\beta}_{t,k}^{\text{F}}=\langle \X_t^{\text{F}}(u) - \overline{\X}^{\text{F}}(u), \phi_k^{\text{F}}(u)\rangle$ represents $k$\textsuperscript{th} principal component score for the female population in year $t$. A similar decomposition can also be implemented for the male series.

Conditioning on the past curves $\bm{\X}(u) = [\bm{\X}_1(u),\dots,\bm{\X}_n(u)]$ for both series and the estimated functional principal components $\bm{\Phi}(u) = \big[\widehat{\bm{\phi}}^{\text{F}}(u), \widehat{\bm{\phi}}^{\text{M}}(u)\big]$, the $h$-step-ahead point forecast of $\bm{\X}_{n+h}(u)$ can be expressed as
\begin{align*}
\widehat{\bm{\X}}_{n+h|n}(u) &= \text{E}[\bm{\X}_{n+h}(u)|\bm{\X}(u),\bm{\Phi}(u)] \\
&= \overline{\bm{\X}}(u) + \bm{\Phi}(u)\widehat{\bm{\beta}}_{n+h|n}^{\top},
\end{align*}
where $\overline{\bm{\X}}(u) = \big[\overline{\X}^{\text{F}}(u), \overline{\X}^{\text{M}}(u)\big]^{\top}$ denotes the mean function aggregated over two populations, and let $\widehat{\bm{\beta}}_{n+h|n}^{\top} = \big[\widehat{\bm{\beta}}_{n+h|n}^{\text{F}}, \widehat{\bm{\beta}}_{n+h|n}^{\text{M}}\big]$ denote the $h$-step-ahead forecasts of the principal component scores obtained from a univariate time-series forecasting method. 

\section{Forecasting the Australian sub-national population}\label{sec:4}

According to Section~\ref{sec:3.1}, a population forecast can be separated to forecast CCR from age~1 to age 100+ based on the HP model and to forecast the infant age group using the ARIMA model. Firstly, the CCR starting from age 1 for females and males can be calculated using the population from 1971 to 2021. The standard functional principal component decomposition method is then used to fit the data, as described by \cite{RD91}. After obtaining the fitted model, we can use the model to forecast CCR for one year ahead, that is 2022. 

\begin{figure}[!htb]
\centering
\subfloat[Female population in Australia]
{\includegraphics[width=8.2cm]{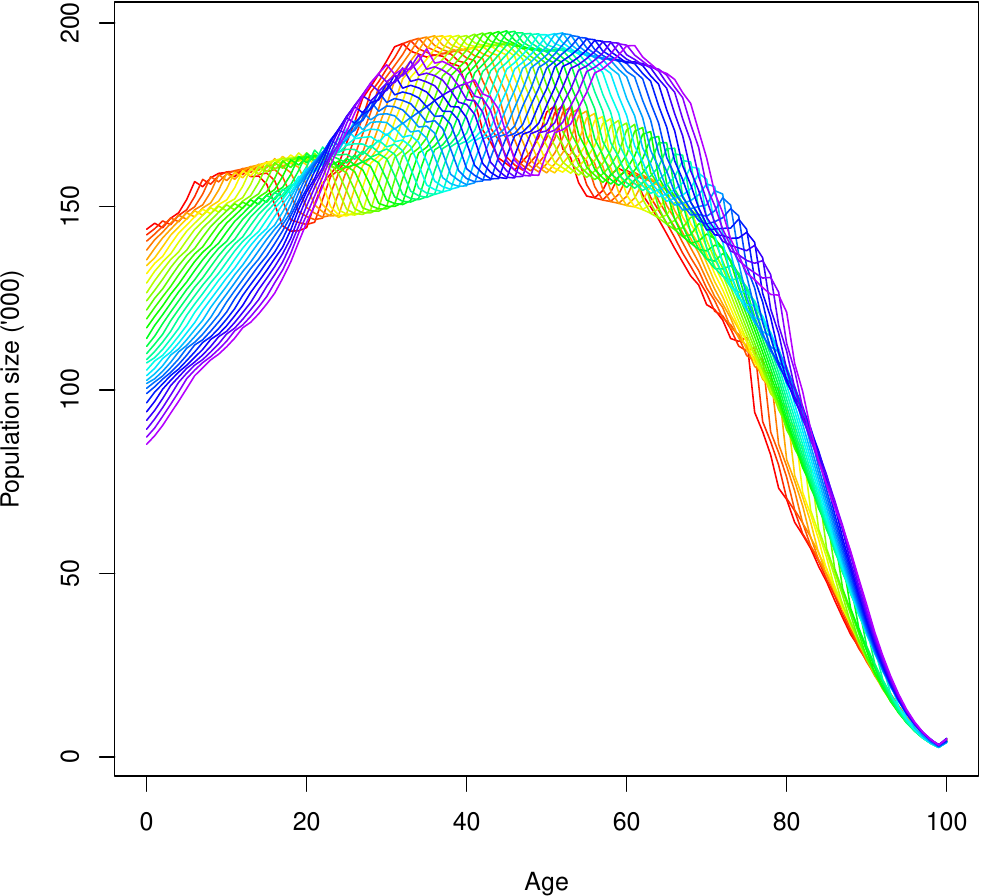}}
\qquad
\subfloat[Male population in Australia]
{\includegraphics[width=8.2cm]{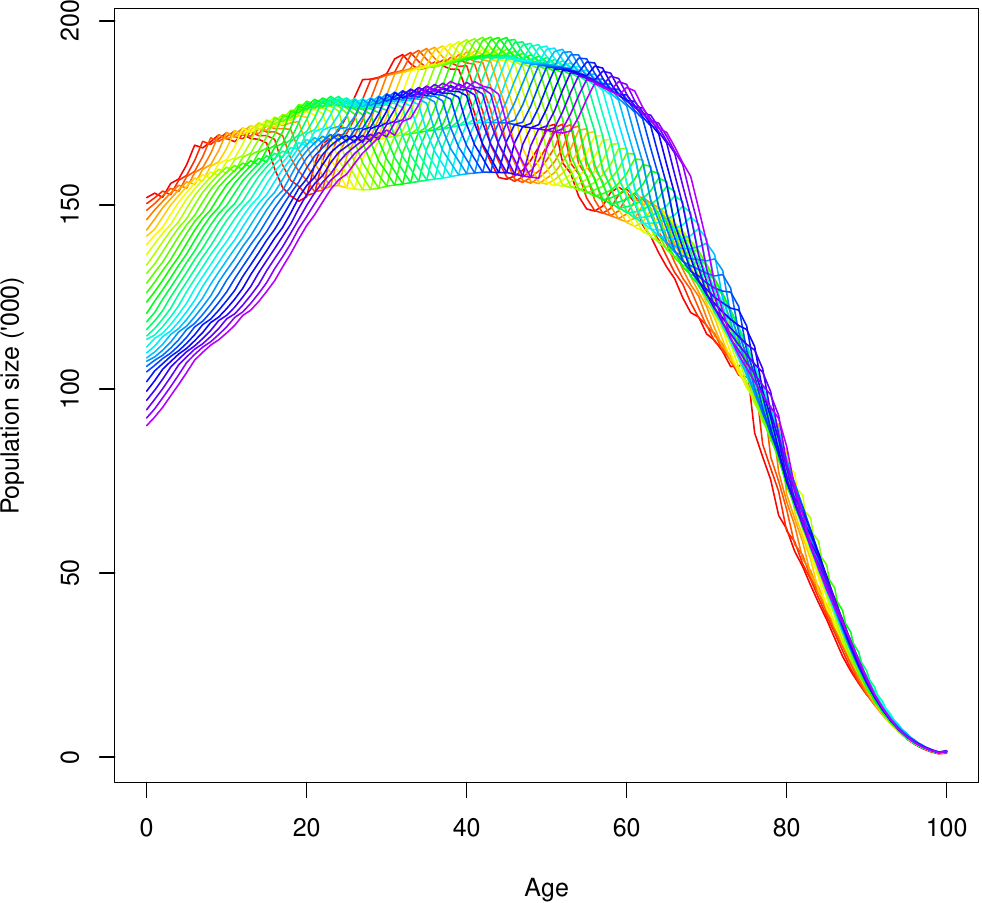}}
\caption{\small{Rainbow plots of Australian female and male population forecasts from 2022 to 2051 between ages 0 and 99 in single years of age, with the last age group being 100+.}}\label{fig:2}
\end{figure}

To obtain the CWR, that is, age 0--1, the ARIMA model is used to model the CWR time series from 1971--2021. Similarly, a one-year ahead forecast for the year 2022 is obtained using automatically selected ARIMA models. Then, the infant age group for males and females can be obtained by~\eqref{eq:sex_ratio}. By combining the 2022 CWR and CCR forecast and multiplying the population in 2021,  we can obtain the population forecast for 2022. Therefore, incorporating the population forecast into the original dataset, the CCR and CWR of 2022 can be calculated accordingly. Incorporating the forecast population into the original population and repeating this process 30 times, we obtain the age-specific population forecast for 2022--2051. By aggregating the sub-national population forecasts, we obtain the population size forecast by sex from 2022 to 2051 in Figure~\ref{fig:2}. It indicates that the trend for a population for different age groups in the next 30 years varies. For example, the infant age group for females is increasing, but the infant age group for males is decreasing. 

The functional time series approach provides a robust population forecast by treating male and female populations as independent variables. However, this method does not account for the correlation between the two sexes, potentially overlooking important dynamics. To address this limitation, we introduce the ``Multivariate Functional Time Series'' (MFTS) method.

The MFTS method enhances forecasting accuracy by capturing the correlation between male and female populations. It does this by combining the Coherent Compositional Reconciliation (CCR) for both sexes, treating them as dependent variables. By incorporating the interdependence between male and female populations, the MFTS method offers a more comprehensive and nuanced forecast. This approach ensures that changes in one sex's population are appropriately reflected in the other, providing a holistic view of population dynamics. As a result, the MFTS method complements univariate functional time series forecasting, delivering a more detailed and accurate prediction of future population trends.

\begin{figure}[!htb]
\centering
\subfloat[Female CCR]
{\includegraphics[width=8.2cm]{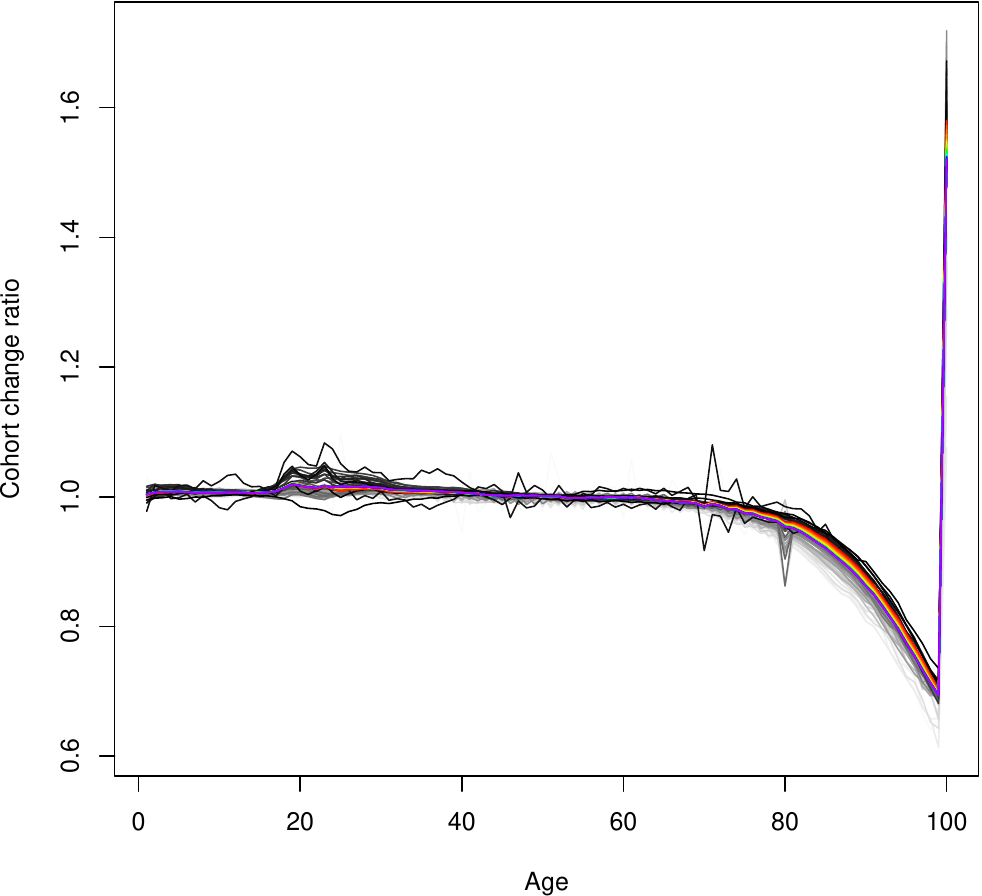}\label{fig:3a}}
\qquad
\subfloat[Male CCR]
{\includegraphics[width=8.2cm]{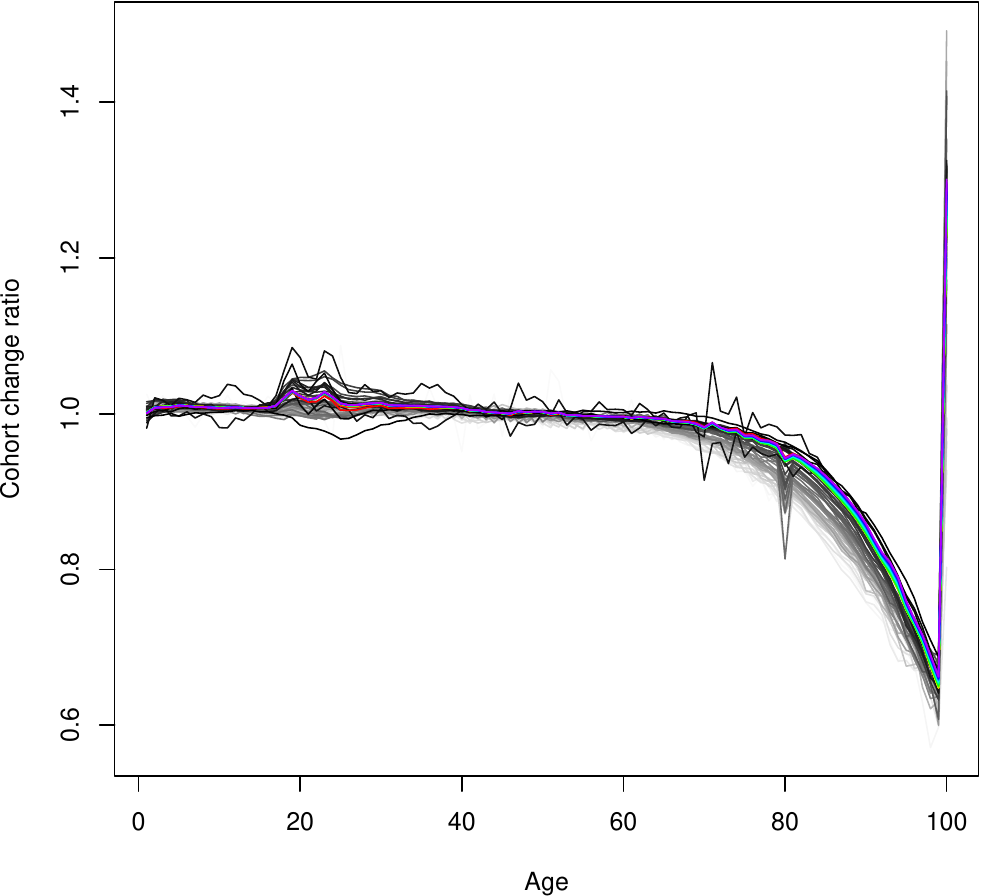}\label{fig:3b}}
\caption{\small{Rainbow plots of Australian female and male CCR from 1972 to 2021 (grey gradient color) and their forecasts from 2022 to 2051 (rainbow color) between ages one and 100+ in single years of age, with the last age group being 100+.}}\label{fig:3}
\end{figure}

Figure~\ref{fig:3} depicts the historical cohort change ratios (CCR) for the Australian female and male populations, respectively, from 1972 to 2021. The graphs use a grey gradient to represent the historical data across the age spectrum from 1 to 100+ years, with the final age group capturing all ages above 100. The projected CCR for the next 30 years (2022-2051) is also shown in a vibrant rainbow color scheme. Upon analyzing these plots, it can be observed that both genders depict a relatively stable crude birth rate (CCR) of around 1.0 for most age groups, indicating a consistent population size from one generation to the next within these cohorts. The colors of the plots transition smoothly across the spectrum, which indicates a continuum of time. The grey gradient blends into the colors of the rainbow while moving from historical data to future forecasts. These visualizations offer valuable insights into long-term demographic trends and can significantly aid resource planning and policy formulation in response to changing population dynamics.

\begin{figure}[!htb]
\centering
\subfloat[Univariate functional time series method]
{\includegraphics[width=8.2cm]{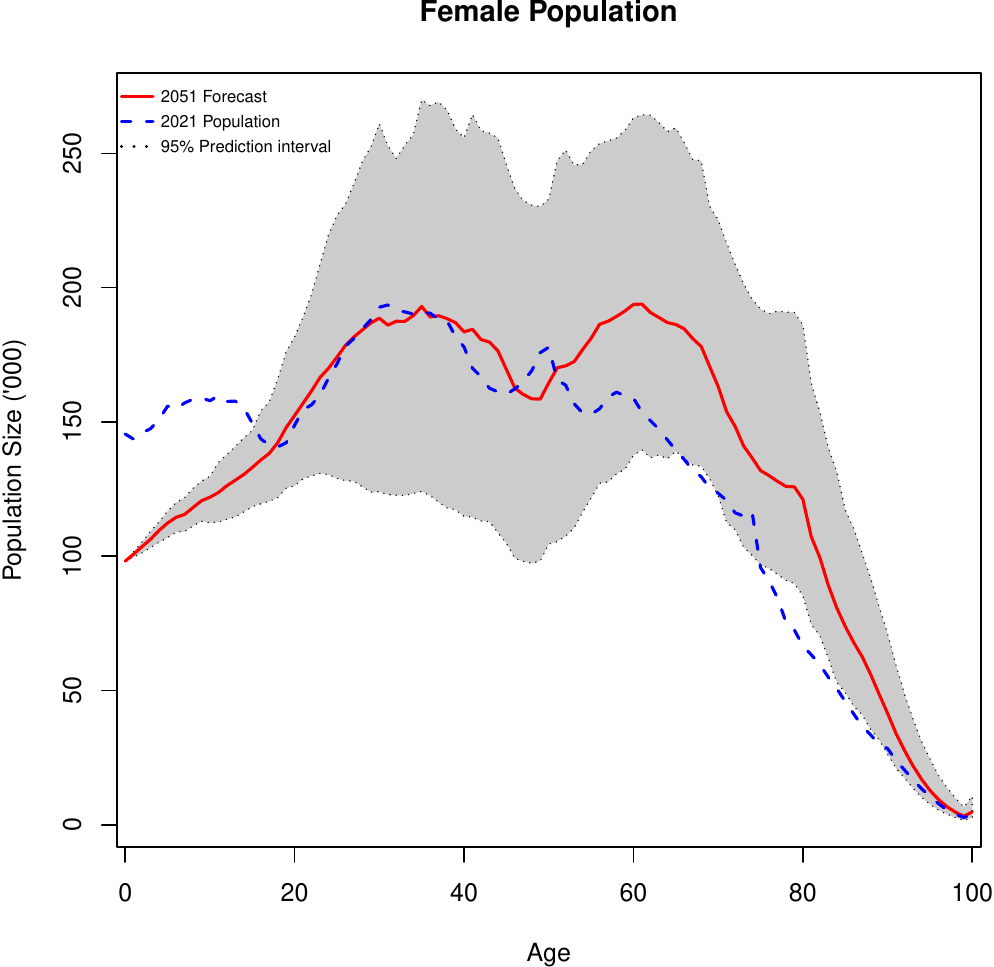}}
\qquad
\subfloat[Univariate functional time series method]
{\includegraphics[width=8.2cm]{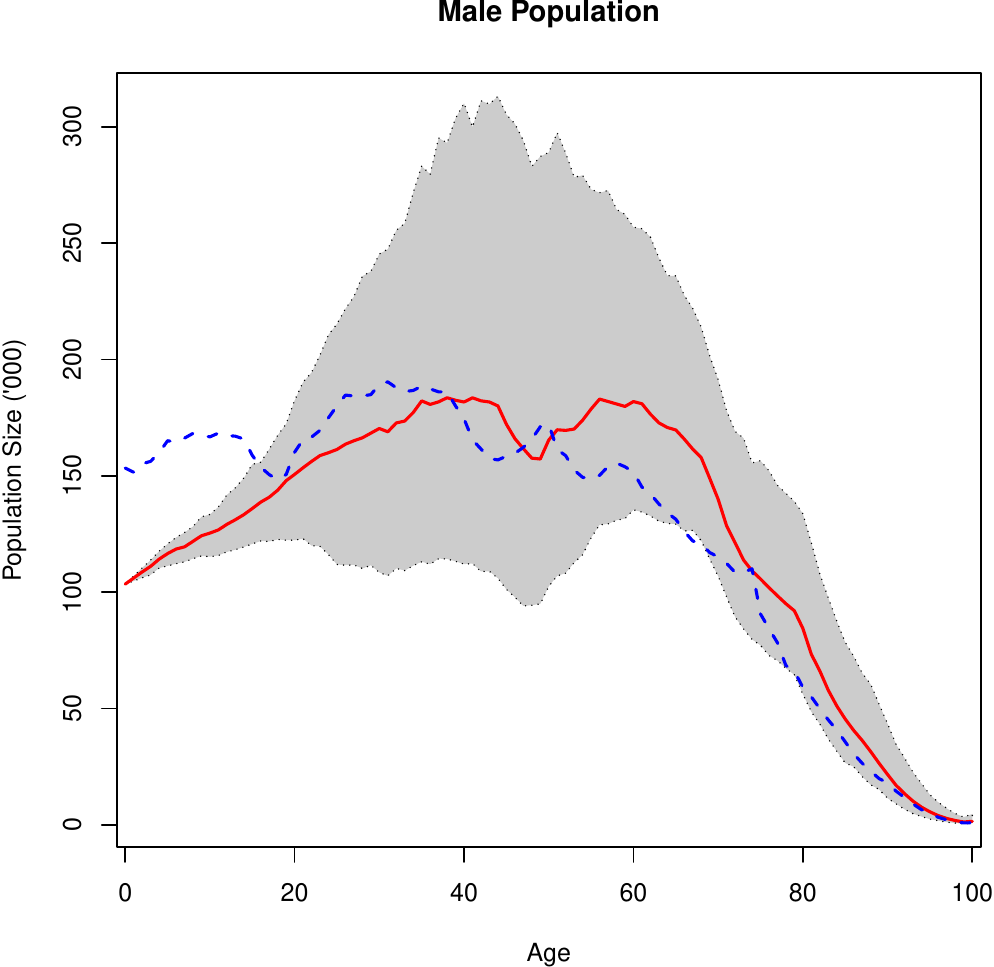}}
\vspace{0.5cm} 
\subfloat[Multivariate functional time series method]
{\includegraphics[width=8.2cm]{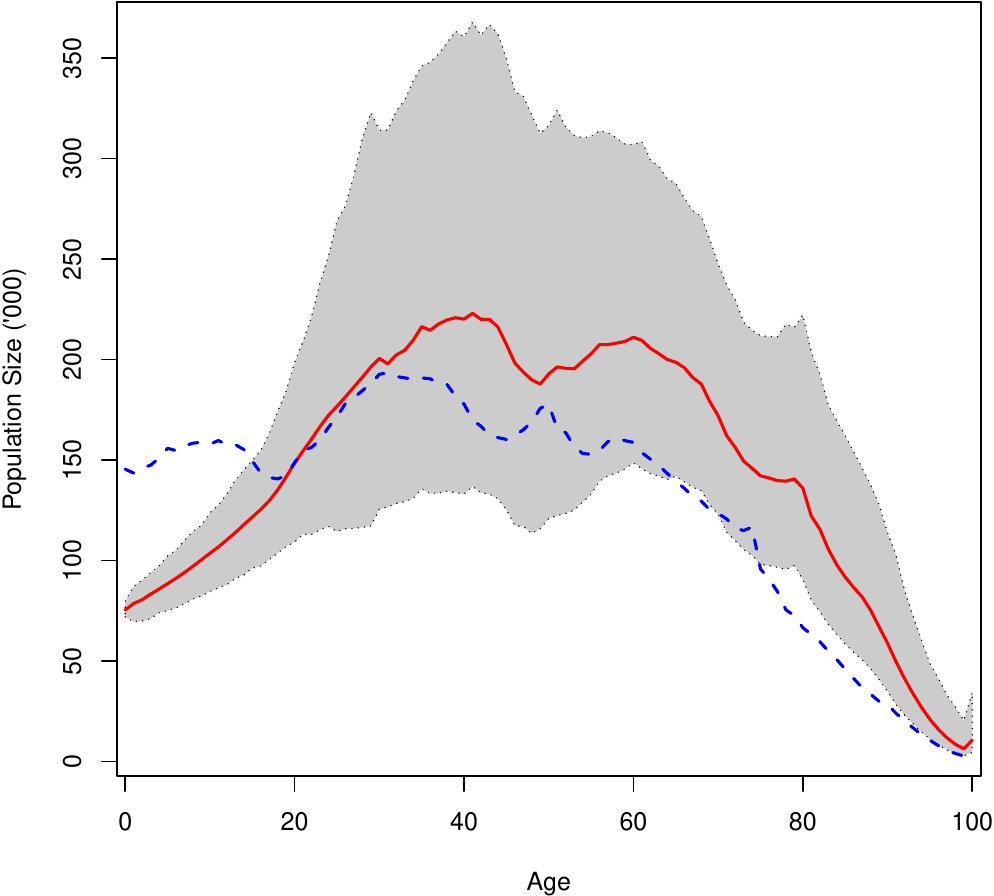}}
\qquad
\subfloat[Multivariate functional time series method]
{\includegraphics[width=8.2cm]{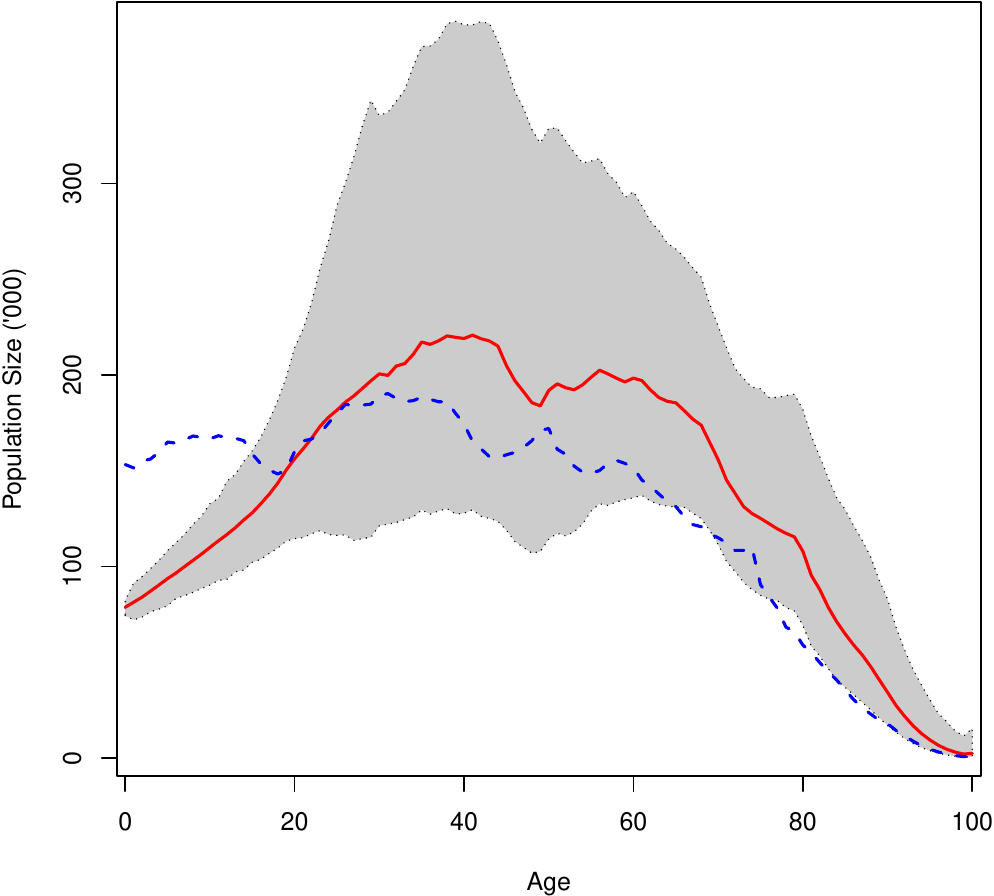}}
\caption{\small{Population for 2021 and its forecast for 2051 for each sex, along with 95\% prediction intervals.}}\label{fig:4}
\end{figure}

Figure~\ref{fig:4} shows a solid red line representing the forecast female and male population for 2051 (a 30-year forecast). In contrast, the blue dashed line represents the population baseline for 2021. The grey area within the figure represents all the possible scenarios within the 95\% pointwise prediction interval for the simulated population paths for 2051. Observing the figure, it is noticeable that both the male and female populations exhibit similar prediction intervals. However, the male population may experience a faster growth rate than the female population. Notably, the prediction intervals for both males and females are more refined in Figures~\ref{fig:4} (c) and (d), demonstrating the MFTS method's capability to account for the correlation between sexes. However, the differences between the univariate and the multivariate functional time series methods are small. Although the MFTS method provides a distinct forecast, the overall influence on future population trends is similar. This suggests that while the MFTS method offers a different perspective by considering the interdependence between male and female populations, the general trends and impacts on future population projections remain consistent across both methods.

\begin{figure}[!htb]
\centering
\subfloat[Univariate functional time series method]
{\includegraphics[width=7.75cm]{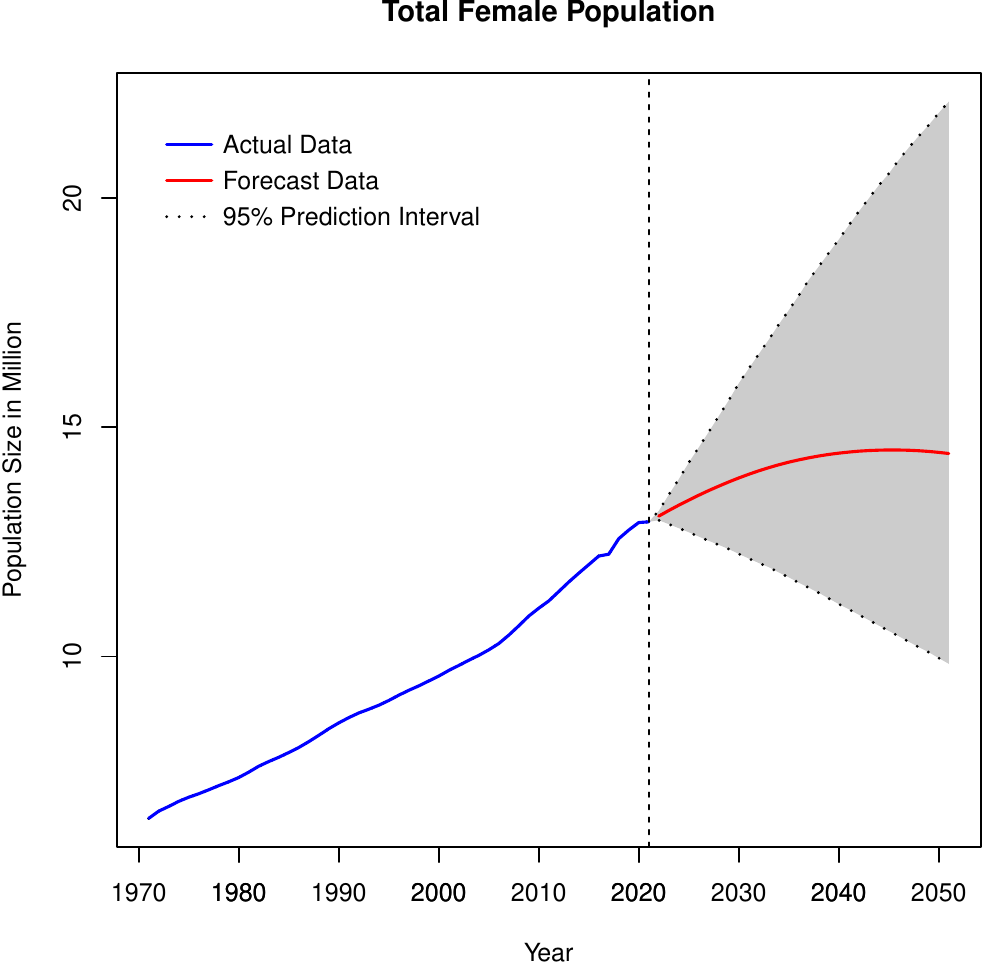}}
\qquad
\subfloat[Univariate functional time series method]
{\includegraphics[width=7.75cm]{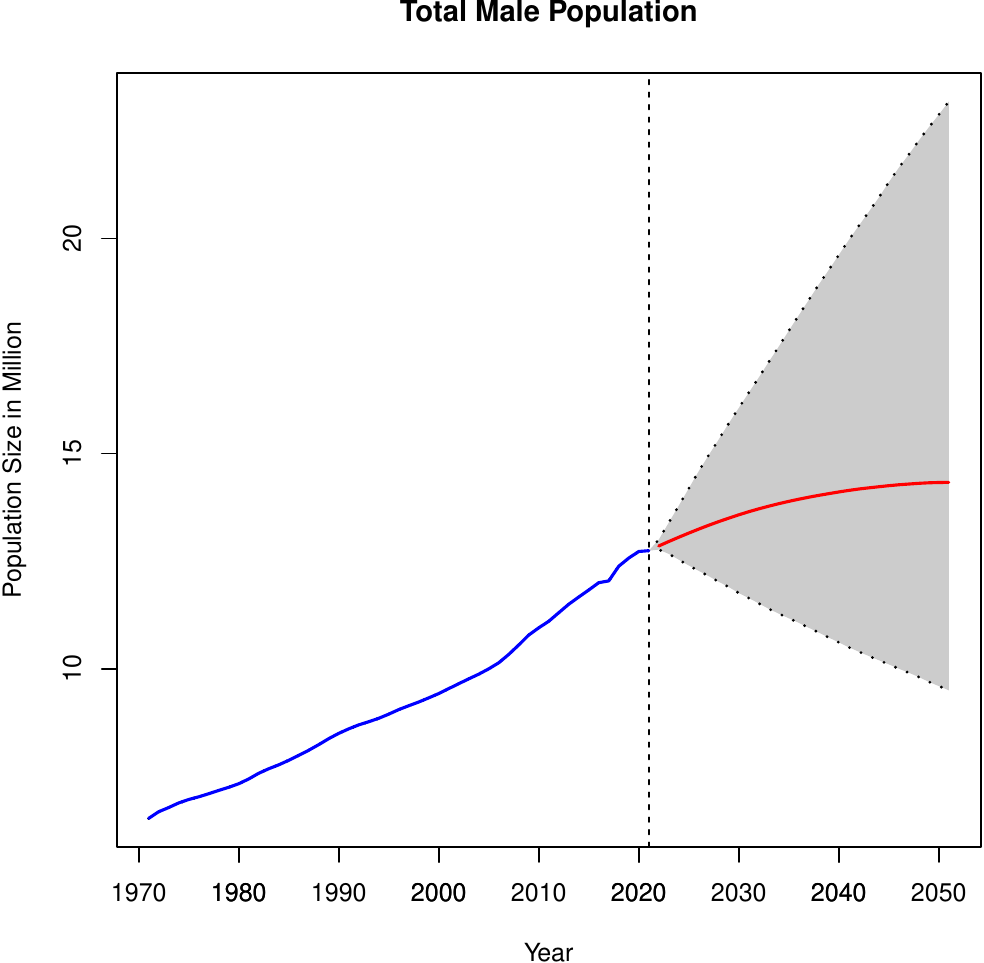}}
\vspace{0.5cm}
\subfloat[Multivariate functional time series method]
{\includegraphics[width=7.75cm]{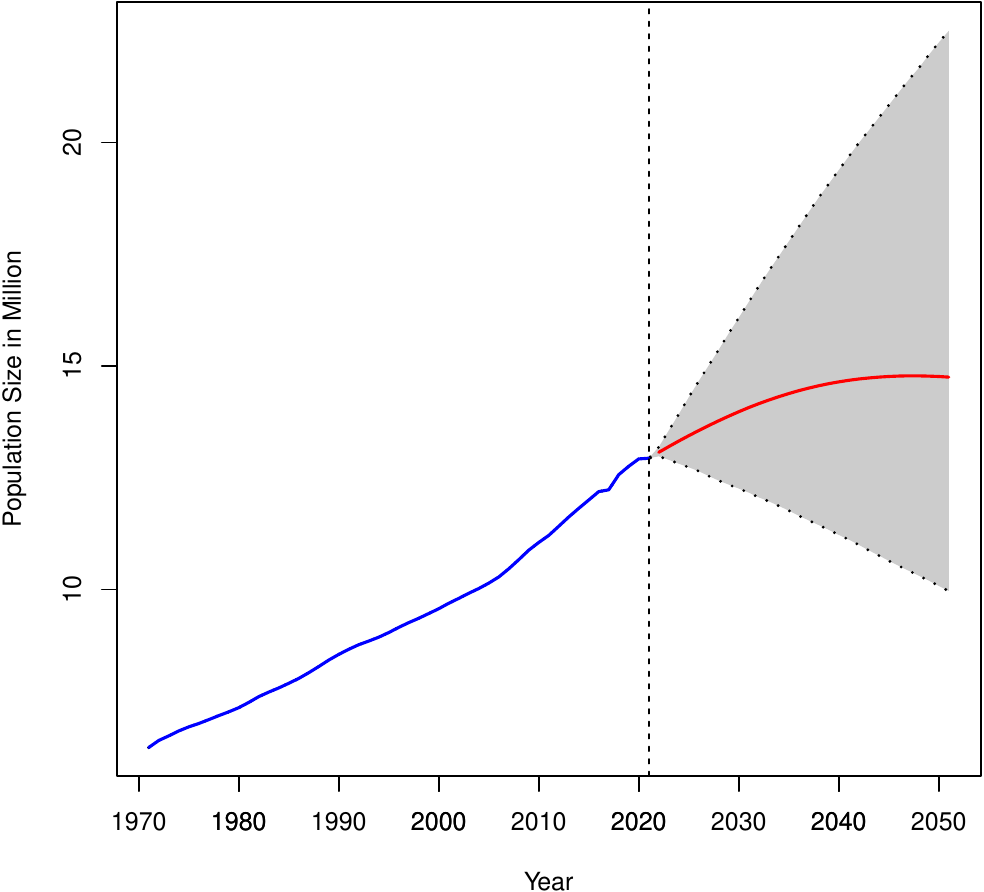}}
\qquad
\subfloat[Multivariate functional time series method]
{\includegraphics[width=7.75cm]{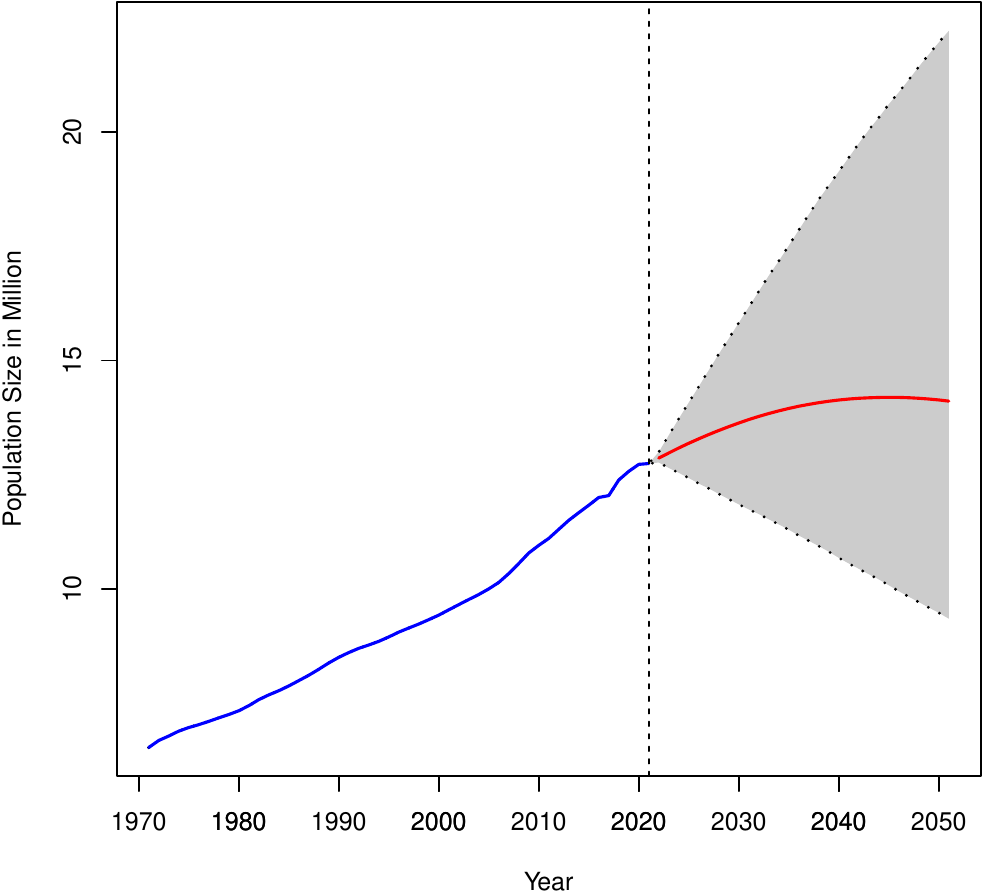}}
\caption{\small{30-year forecasts of the total population for each sex, along with 95\% prediction intervals.}}\label{fig:5}
\end{figure}

Figure~\ref{fig:5} displays the forecast total population for males and females. The line depicted in the graph illustrates the current total population for both genders. In contrast, the dashed line within the grey area represents the estimated total population from 2022 to 2051. Furthermore, the grey area represents all possible results within 95\% prediction intervals for the total population. Figures~\ref{fig:5} (c) and (d) present the forecasts using the MFTS method. The blue line represents actual population data up to 2021, while the red line shows the forecasted population from 2022 to 2051. Notably, the point forecasts using the MFTS method are slightly lower than those obtained using the functional time series method. Additionally, the grey-shaded areas representing the 95\% prediction intervals are slightly wider in the MFTS forecasts. 

This suggests that while the MFTS method accounts for more uncertainty, it provides a comprehensive view by including the correlation between male and female populations, potentially offering a more nuanced understanding of future population trends. Overall, Figure~\ref{fig:5} highlights the differences and similarities between the two forecasting methods, demonstrating that although the point forecasts differ slightly, the general trends and prediction intervals remain broadly consistent. This comparison underscores the value of considering multiple methods to capture a range of potential future scenarios.

\begin{figure}[!htb]
\centering
\includegraphics[width=9cm]{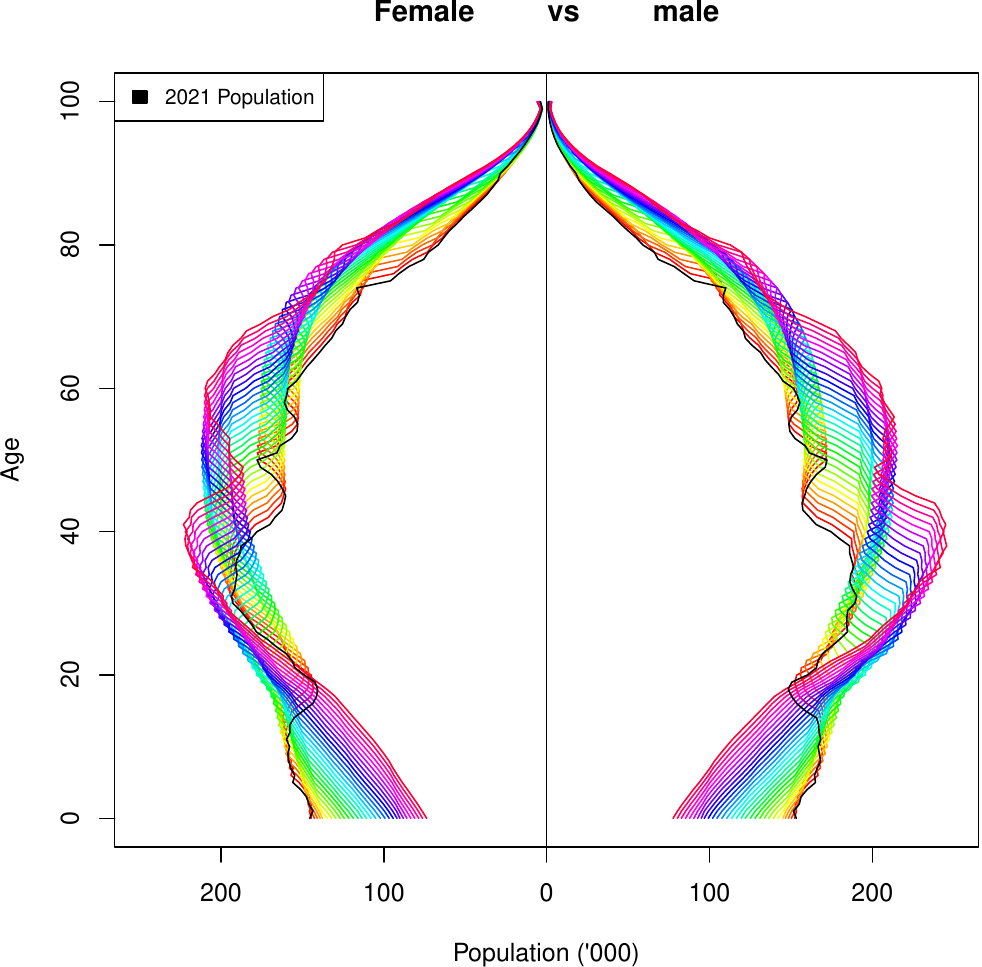}
\caption{\small{Holdout and forecast population pyramids of males and females. The black line represents the population pyramid for females and males in 2021. The rainbow lines represent simulated paths of the forecast population of females and males from 2022 to 2051.}}\label{fig:6}
\end{figure}

Figure~\ref{fig:6} presents the population pyramids of the forecast population from 2022 to 2051, produced by the functional time series forecast method, with the population in 2021 as a baseline. Figures~\ref{fig:5} and~\ref{fig:6} indicate that there will be a significant rise in the number of males and females in the next 30 years. The male population is expected to grow more rapidly than the female population during this period. Population growth is a crucial factor that the Federal Government needs to track to ensure that the community is prepared and equipped with the necessary resources. Forecasting is vital to making informed decisions and taking preventive measures to meet the future needs of the growing population.
 
\begin{figure}[!htb]
\centering
\subfloat[Total female population]
{\includegraphics[width=8.55cm]{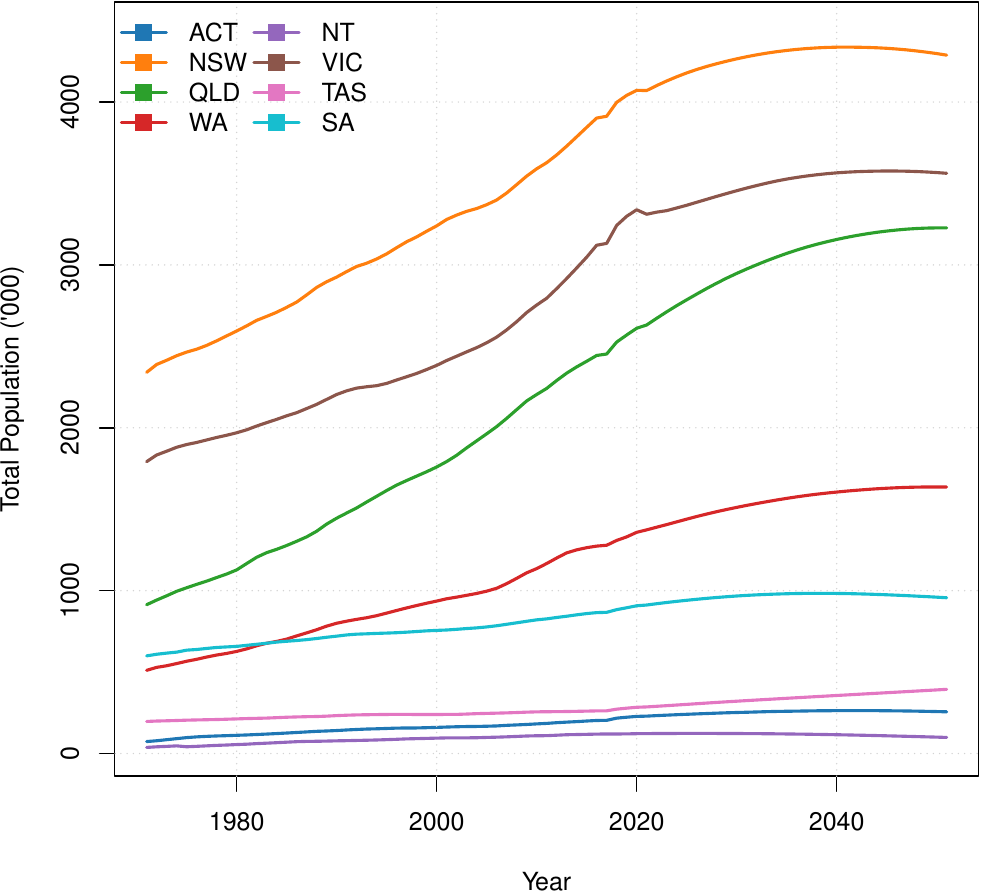}\label{fig:7a}}
\qquad
\subfloat[Total male population]
{\includegraphics[width=8.55cm]{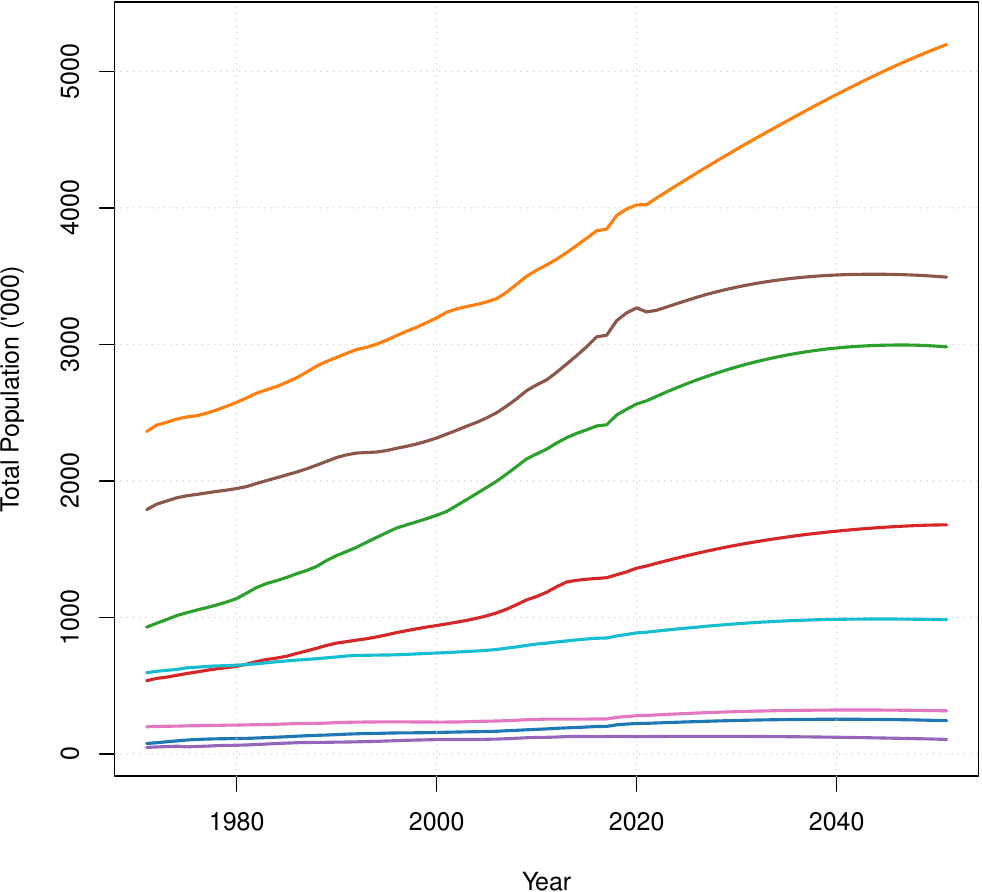}\label{fig:7b}}
\caption{\small{Total population for each state for different gender.}}\label{fig:7}
\end{figure}

Figure~\ref{fig:7} presents a comparative graph analysis of the total population trends by gender for various Australian states and territories from 1972 to 2051. Figures~\ref{fig:7a} and~\ref{fig:7b} delineate the population trajectories for females and males, respectively. Each line within the graph corresponds to a different state, such as ACT, NSW, VIC, and so on.

The data portrayed in the graphs indicate a continuous demographic expansion across all jurisdictions. This upward trajectory is consistent with general population growth trends, although the gradient of each line varies, reflecting differing growth rates among the states and territories. For instance, the trajectory for NSW suggests a more pronounced growth rate, which could indicate a higher birth rate, lower mortality rate, increased migration, or a combination of these factors.

No abrupt discontinuities are evident within the presented data range, suggesting a steady population increase rather than sudden demographic shifts. These trends have significant implications for policy planning, particularly in determining sustainable pension ages. The gradual population increase, especially within the working-age cohort, will directly affect the pension system, necessitating adjustments to pension ages and related social security frameworks to ensure long-term sustainability.

\begin{figure}[!htb]
\centering
\includegraphics[width=9.23cm]{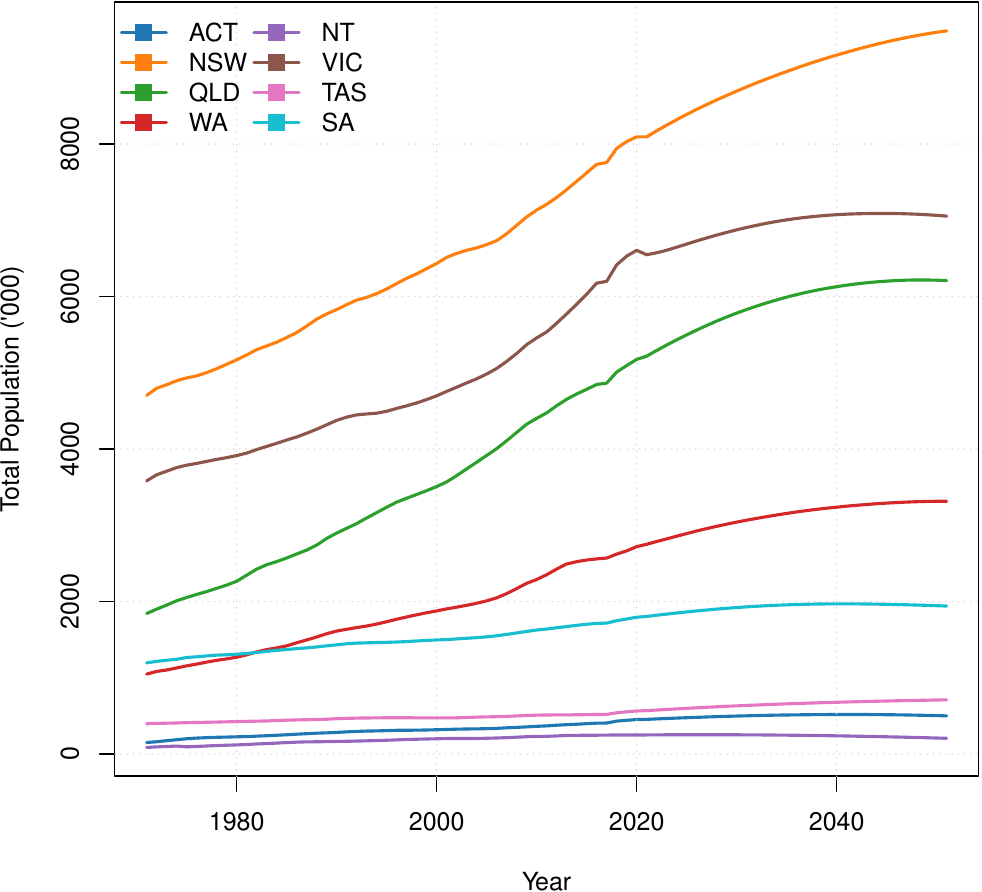}
\caption{\small{Total population for each state.}}\label{fig:8}
\end{figure}

Figure~\ref{fig:8} exhibits a graphical display representing each Australian state's total population across a temporal span from 1972 to 2051. The graph is not segmented by gender, and it provides an aggregate overview of demographic trends within each region. Similarly to Figure~\ref{fig:7}, each line on the graph represents a different state or territory, with the same labels from Figure~\ref{fig:7} serving as identifiers. The lines depict a clear upward progression in population numbers, indicating overall growth within these regions. The slope of each line varies, suggesting that the rate of population increase is not uniform across the states and territories. Some states, such as NSW and VIC, are typically populous. They display a steeper incline, indicative of a more rapid population rise.

\begin{figure}[!htb]
\centering
\includegraphics[width=9.23cm]{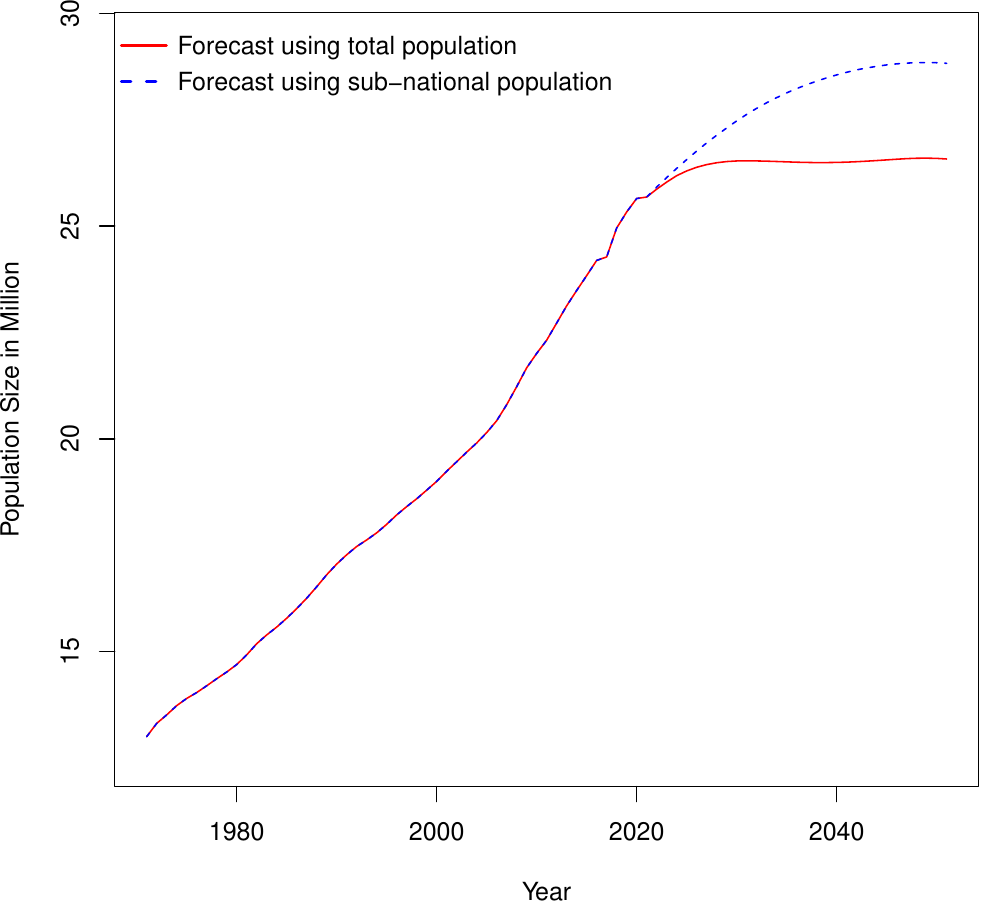}
\caption{\small{Comparison of national vs. sub-national population forecasts for Australia (2022-2051).}}\label{fig:8.5}
\end{figure}

In Figure~\ref{fig:8.5}, we display the population forecasts for Australia and the combination of forecasts from its subnational level. Two different population forecasting models were fitted to the population data from 1971 to 2021 to produce forecasts from 2022 to 2051. The forecasting model based on the Australian national data can smooth out regional variations and anomalies, providing a broader but less detailed perspective. On the other hand, the forecasting model using subnational population data considers regional specification, capturing local trends and demographic changes, thus offering more accurate and detailed predictions.

From Figure~\ref{fig:8.5}, it reveals that the two forecast lines may diverge, indicating differences in prediction when using aggregated national data versus detailed subnational data. The subnational forecasts in red show high variability, emphasizing the importance of considering regional differences in making population forecasts. The divergence of the lines over time can indicate changing regional dynamics that might not be apparent when studying at national data alone.

\section{Determining sustainable pension age}\label{sec:5}

As a caveat of age-specific population forecasting, we investigate the effect of population changes on the pension age. Australia's aging population is a result of three factors:
\begin{inparaenum}
\item[(1)] significant technological advances in medical care leading to improvements in living conditions and longer life expectancy;
\item[(2)] high fertility rates after World War II resulting in the ``baby-boomer" generation between the mid-1940s and the early 1960s, with people born during this period now entering, if not having already entered, older age groups, contributing to the overall aging of the population; and
\item[(3)] a large decline in fertility rates over the past 30 years \citep{FJK08}, which means that when fertility rates are below the replacement level (the number of children needed to replace the parents), the population tends to age because there are fewer births relative to the number of older individuals.
\end{inparaenum}
As a result, many older people are in retirement, and relatively fewer workers are in the workforce. 

Many other developed countries face similar demographic challenges. One way of measuring the problem is using the old age dependency ratio (OADR), expressed as
\begin{equation*}
\text{OADR} = \frac{\text{number of people aged over pension age}}{\text{number of people aged 15 to pension age}}\times 100\%.
\end{equation*}
To measure the financial burden on the Australian workforce, the OADR is a demographic indicator we use to determine a sustainable pension age. A rise in the OADR will reduce the ability of the working-age population to finance pensions and health costs \citep{ARW07}. 

We provide forecasts of future population age and sex structures over the next 30 years from 2022 to 2051 and use those forecasts to compute the OADR, resulting in a sustainable pension scheme. We find a sustainable pension age leading to a specified OADR value. Following the early work by \cite{HZS21}, the specified OADR value has been set at 23\%, which we follow. 

We aim to determine a pension age scheme that would lead to the OADR being below the desired threshold, thus imposing a sustainability constraint. Let $a_t$ denote the pension age in year $t$, and we denote a pension age scheme over the prediction horizon by $\boldsymbol{P}=\left[a_{T+1}, \ldots, a_{T+H}\right]^{\top}$, with the corresponding OADR values given OADR lower than the desired OADR threshold. We also assume that the adjustment unit of the pension age is one year for easier implementation.

According to \cite{HZS21}, let $O^*$ denote the desired OADR threshold. Because we do not observe the population in future years, we must estimate $O_{T+h}$ from the simulated populations. Then, we compute the pension age scheme $\boldsymbol{P}$ using the following algorithm. Starting with $h=1$:
\begin{itemize}
\item set $a_{T+h}=a_{T+h-1}$
\item increment $a_{T+h}$ by a small increment, such as 1-month intervals, until $O_{T+h \mid T} \leq O^*$.
\end{itemize}

We not only determine the target pension age based on the point prediction of OADR, but we also identify ``plausible" pension age schemes that could achieve the desired OADR level. Our search includes a range of pension age schemes where $O^*$ falls within the 95\% prediction intervals of the simulated $O_{T+H \mid T}$. Since the relationship between pension age and OADR is monotonic, we only need to find plausible upper and lower boundaries for pension age schemes. We achieve this by identifying the upper or lower limit of the 95\% prediction intervals of OADR that is equal to the desired OADR level. Any pension age scheme within these boundaries is considered plausible. This range of plausible pension age schemes represents 95\% prediction intervals of pension age schemes that we are confident will yield the desired outcome. The algorithm to find the upper and lower boundary of plausible pension age schemes is the same as above, but the mean of $O_{T+H \mid T}$ is replaced by the 2.5\% and 97.5\% quantiles. 

\begin{figure}[!htb]
\centering
\subfloat[Total population via the univariate functional time series method]
{\includegraphics[width=8.5cm]{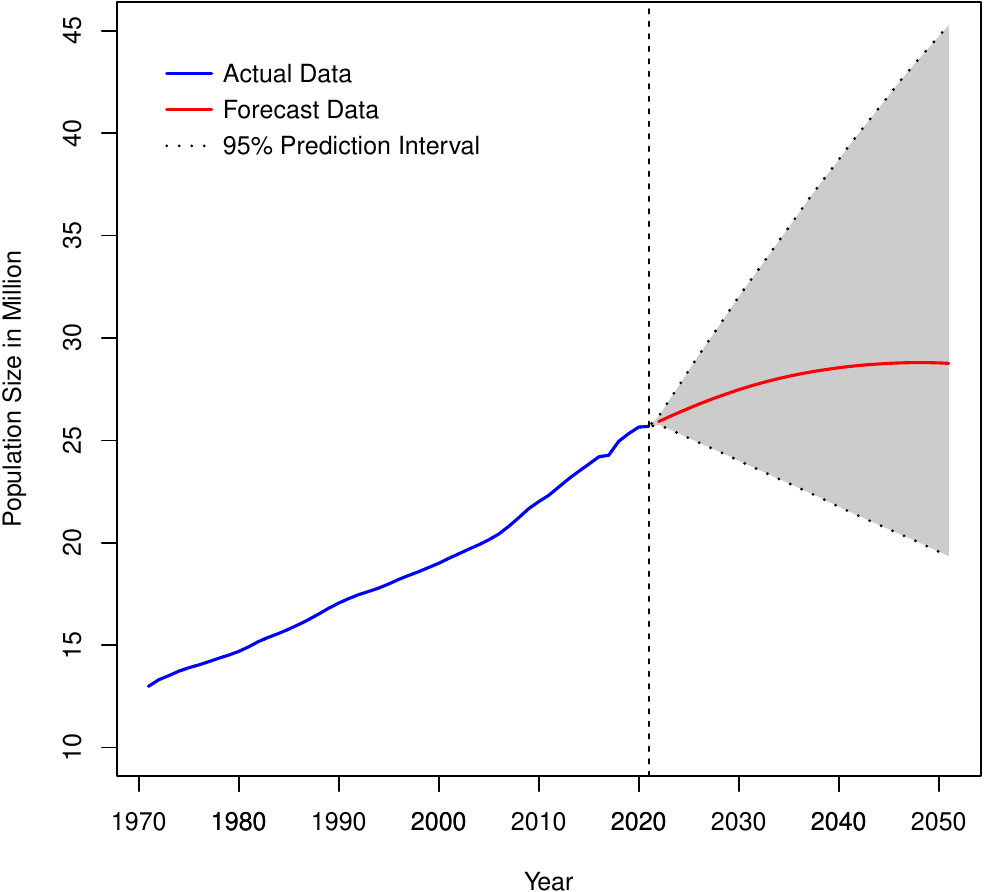}}
\qquad
\subfloat[Total population via the multivariate functional time series method]
{\includegraphics[width=8.5cm]{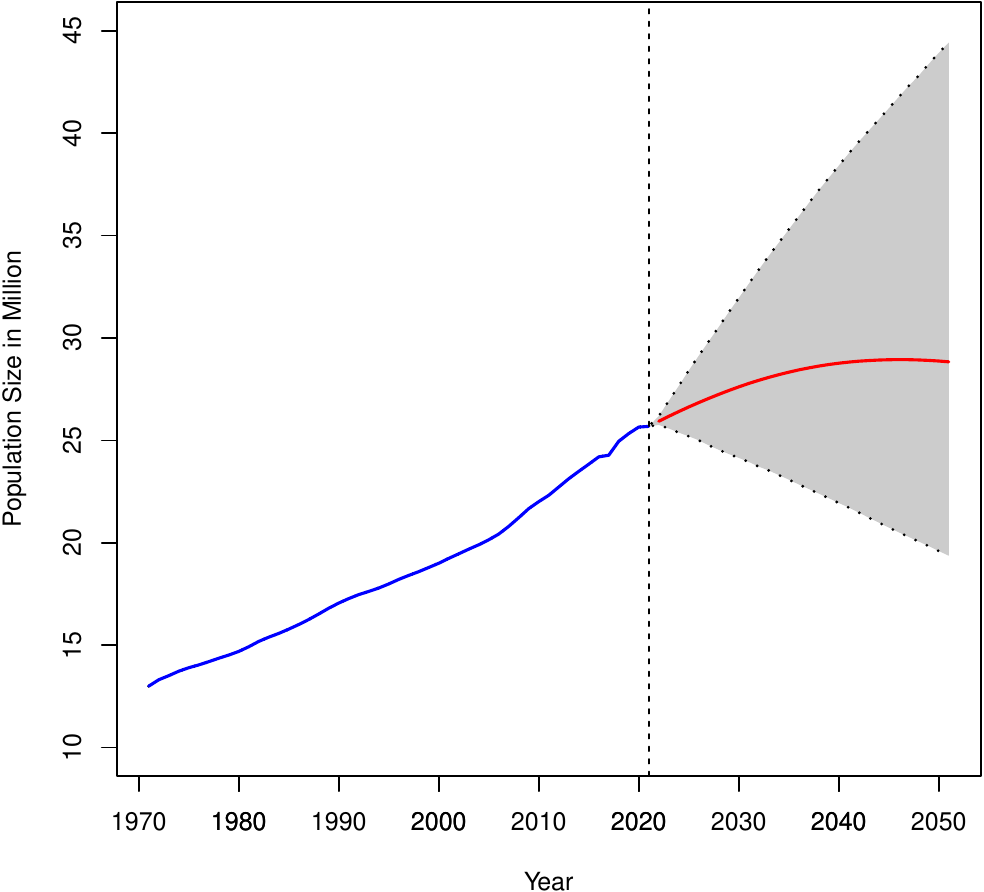}}
\caption{\small{30-year forecasts of the total population, along with 95\% pointwise prediction intervals.}}\label{fig:9}
\end{figure}

Figure~\ref{fig:9} shows the anticipated population of Australia from 2022 to 2051. The solid line indicates the current population, while the dashed line within the shaded area shows the point forecast population for 2022 to 2051. The grey area represents the 95\% prediction interval for the total population. Under the scenario where the population increases rapidly, the target pension age to maintain sustainability will need to be increased. Figure~\ref{fig:9} (b) presents the forecast using the MFTS method. Similar to Figure~\ref{fig:9} (a), the blue line represents actual population data up to 2021, while the red line shows the forecasted population from 2022 to 2051. The grey shaded area indicates the 95\% prediction intervals. Notably, the point forecasts using the MFTS method are slightly lower compared to the univariate method, and the prediction intervals are wider. This indicates that the MFTS method accounts for additional uncertainty by capturing the correlations between male and female populations, which may lead to a more cautious projection.

The overall trend in both figures suggests that the population of Australia will continue to grow, although the rate of growth and the degree of uncertainty vary between the two methods. The MFTS method provides a more comprehensive forecast by incorporating the interdependencies between different demographic groups, potentially leading to more robust planning and policy decisions. For instance, under scenarios of rapid population increase, the target pension age necessary to maintain sustainability may need to be adjusted upwards. The MFTS method's broader prediction intervals highlight the importance of preparing for a range of future outcomes, ensuring that policy measures remain effective under varying demographic conditions.

\begin{figure}[!htb]
\centering
\subfloat[Pension age scheme with the female and male population forecasts obtained from the univariate functional time series method]
{\includegraphics[width=8.5cm]{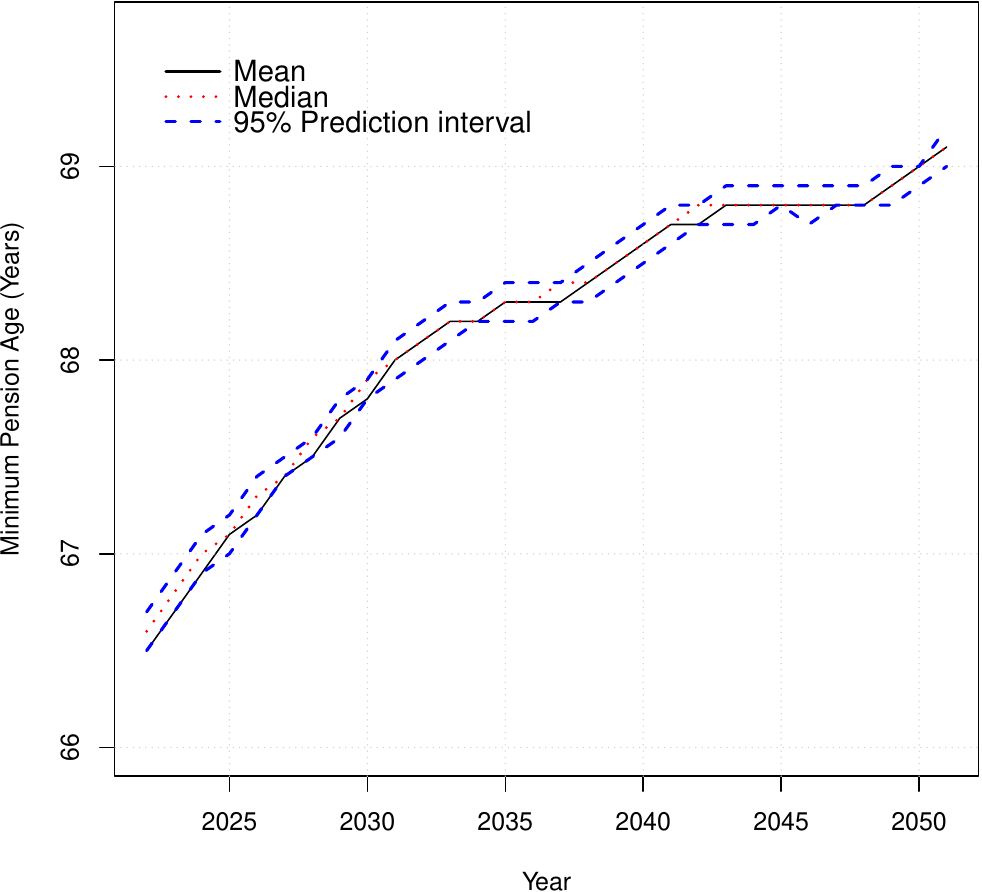}}
\qquad
\subfloat[Pension age scheme with the female and male population forecasts obtained from the multivariate functional time series method]
{\includegraphics[width=8.5cm]{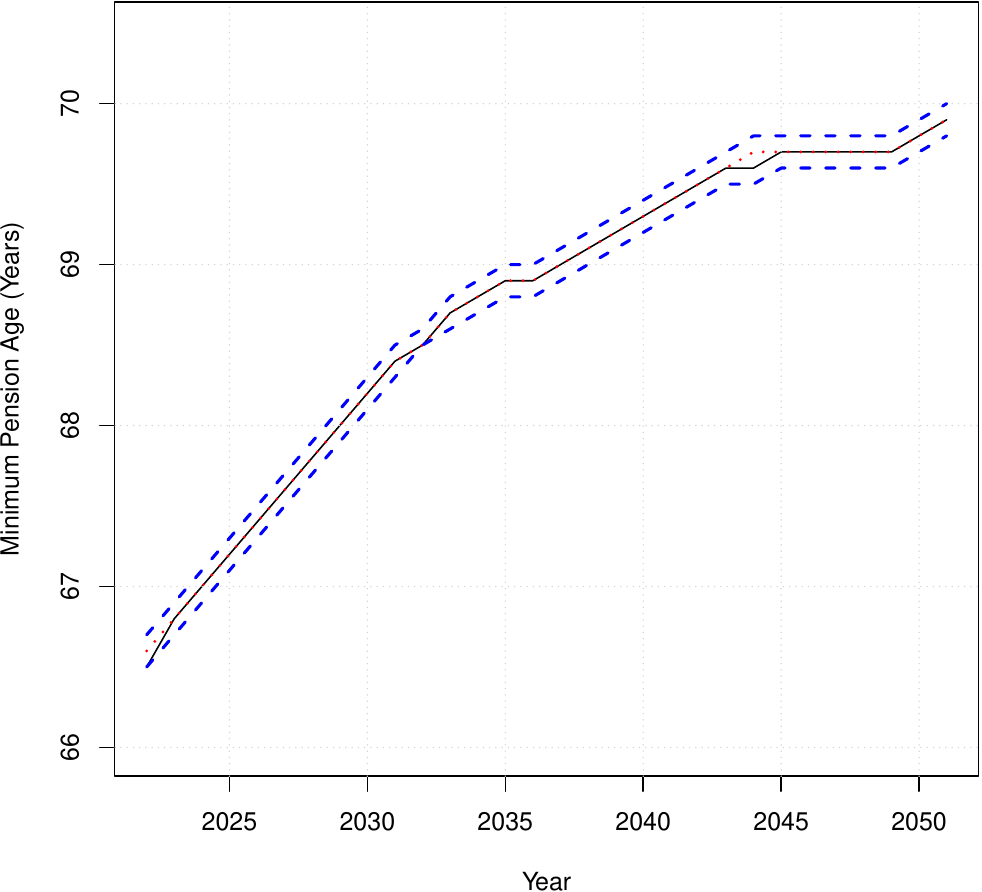}}
\caption{\small{The 30-year forecasts of the pension age scheme with 95\% pointwise prediction intervals.}}\label{fig:10}
\end{figure}

Figure~\ref{fig:10} represents a projection graph that provides a 30-year forecast of the pension age scheme, accompanied by 95\% prediction intervals. The primary focus of Figure~\ref{fig:10} is delineating the minimum pension age from approximately 2022 to 2051. Central to the figure are two key trend lines: the mean, represented by a black solid line, and the median, illustrated by a red dashed line. Surrounding these central trend lines are blue dashed lines depicting the 95\% prediction intervals, offering a probabilistic range within which future values are expected to fall. The trend lines exhibit an upward trajectory, suggesting that the pension age is anticipated to increase over the forecast period. The prediction intervals denote the uncertainty inherent in such long-term forecasts, reflecting potential variability due to economic, demographic, or policy changes. The gradual rise in pension age encapsulated in this figure underscores the demographic and fiscal challenges anticipated to influence pension policy. Figure~\ref{fig:10} (b) illustrates the pension age scheme forecast using the MFTS method. Similar to Figure~\ref{fig:10} (a), the black solid line represents the mean forecast, the red dashed line indicates the median forecast, and the blue dashed lines show the 95\% prediction intervals. Notably, the point forecasts for the MFTS method are slightly higher than those obtained using the univariate method, and the prediction intervals are marginally wider. This indicates that the MFTS method captures additional uncertainty by accounting for the correlations between male and female populations, resulting in a more conservative projection.

Figure~\ref{fig:10} emphasizes this forecast's critical nature in ensuring the sustainability of the Australian pension system, suggesting that policy adjustments (e.g., raising the pension age) may be necessary to maintain the scheme's long-term viability. Based on this visualization, we can infer that the current pension age of 67 will remain sustainable until 2025. However, if the Australian Government wants the scheme to remain sustainable until around 2051, it must increase the pension age to 69. In the worst-case scenario, raising the pension age to 69.5 before 2051 will be necessary to ensure sustainability. It is essential to note that any increase in the target pension age must be done carefully and thoughtfully. The overall trend in both subfigures of Figure~\ref{fig:10} indicates that the minimum pension age will need to rise to maintain the sustainability of the pension system. According to the MFTS method, the optimal pension age in 2051 is projected to be 69.9 years, compared to 69.2 years with the univariate method. Under the worst-case scenario forecasted by the MFTS method, the pension age may need to increase to 70 years to ensure sustainability, compared to 69.5 years with the univariate method. 

These forecasts underscore the importance of policy adjustments to address demographic changes and fiscal challenges. Ensuring the long-term viability of the pension system will require careful consideration of these forecasts and the potential need for gradual increases in the pension age. This visualization highlights the critical nature of these adjustments, emphasizing the need for thoughtful and deliberate policy measures to sustain the pension system for future generations.

\section{Pension welfare distribution} \label{sec:6}

We have shown that the minimum pension age in Australia is expected to increase due to the aging population. To gain more insight into welfare distribution in changing population demographics, we examine age pension income among retirees across all eight states and territories throughout their retirement. 

\begin{figure}[!htb]
\centering
\includegraphics[width = 0.485\linewidth]{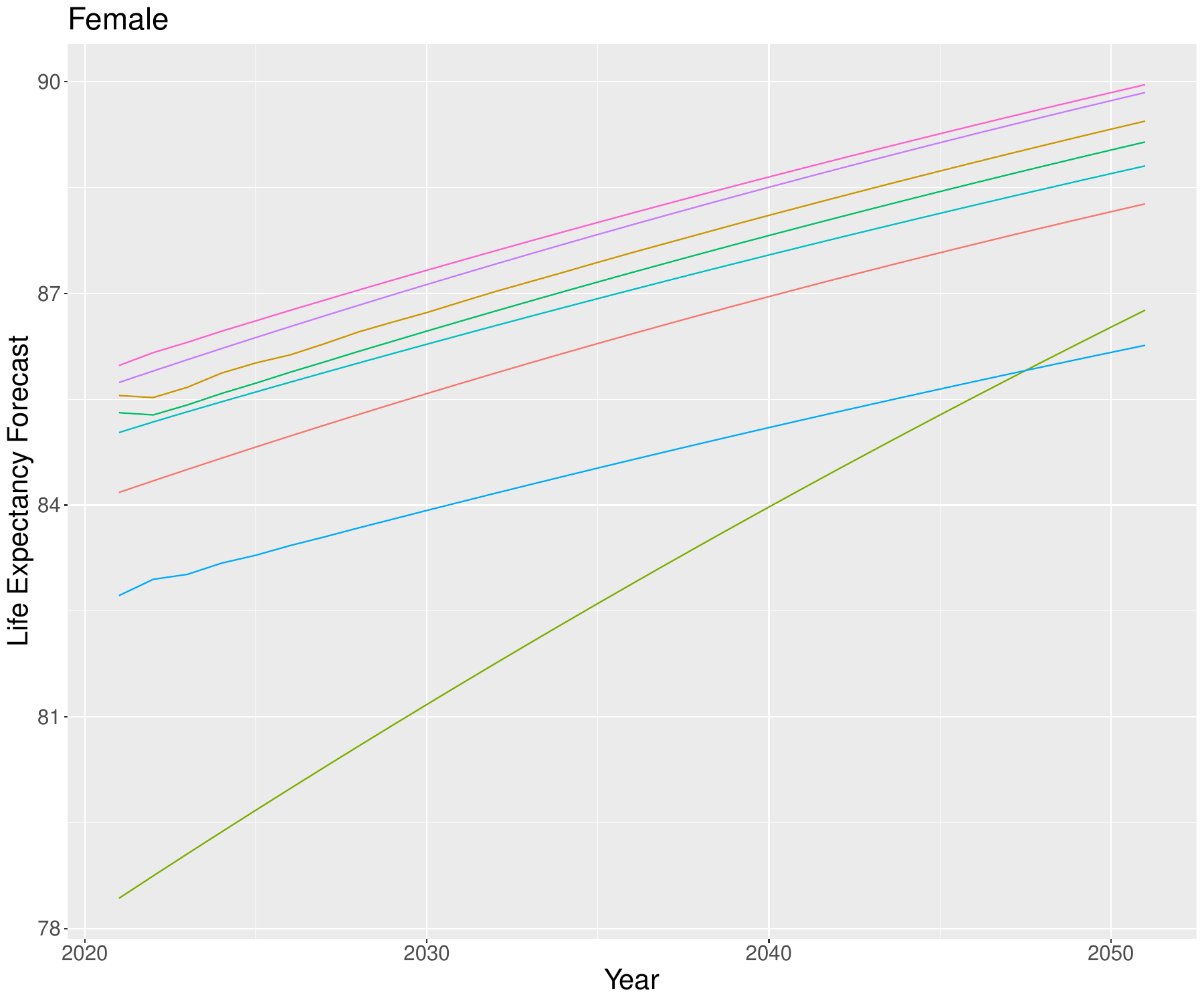} 
\quad
\includegraphics[width = 0.485\linewidth]{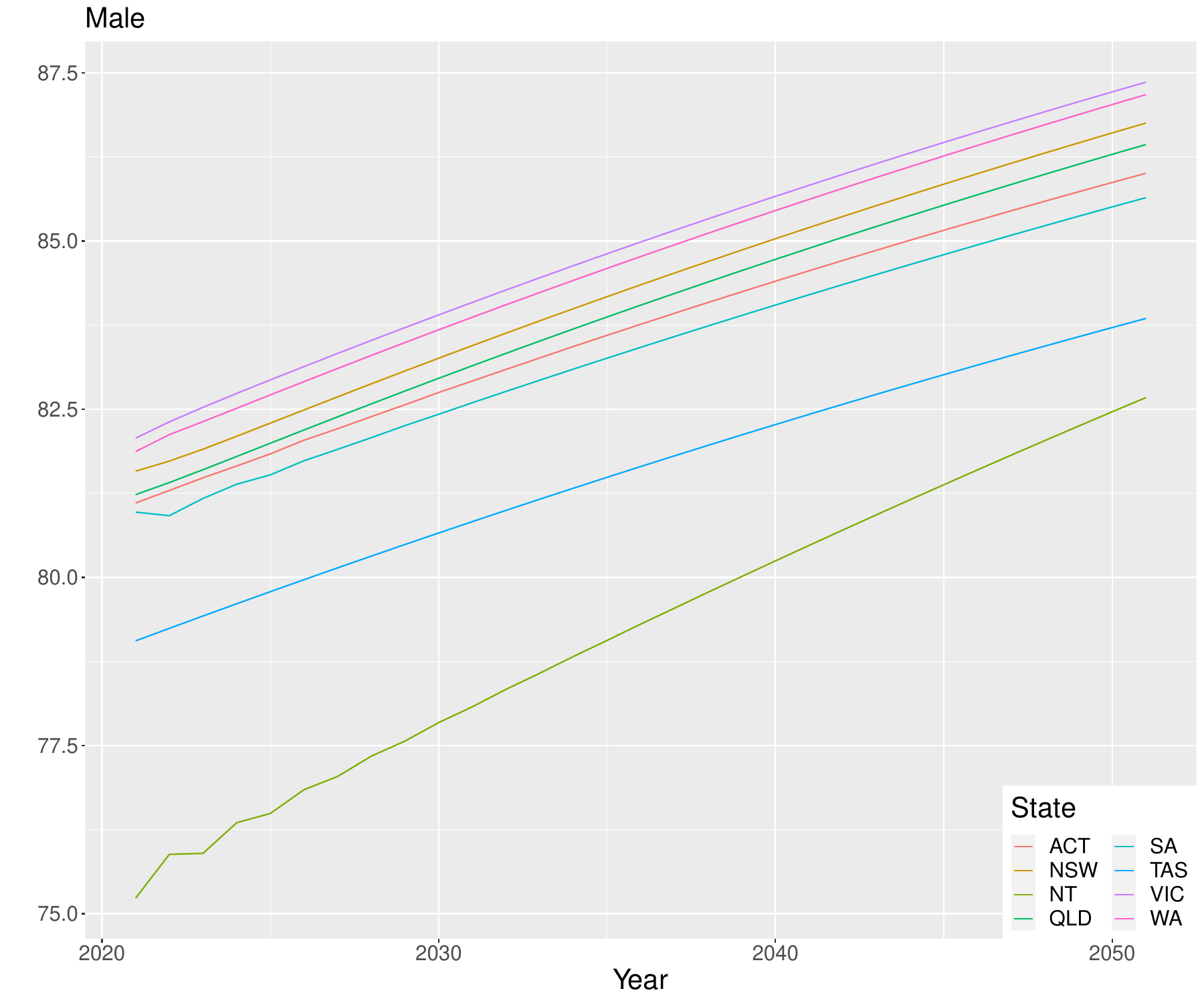} 
\caption{\small{Forecast Life expectancies for residents in all states and territories of Australia.}}\label{fig:11}
\end{figure}

We first use the sub-national age-specific mortality rates obtained from \cite{AHMD23} and the functional time series method to forecast life expectancies for Australians over the period between 2022 and 2051. Figure~\ref{fig:11} indicates that all states and territories of Australia are expected to witness steady increases in residents' life expectancies.

\begin{figure}[!htb]
    \centering
    \includegraphics[width=0.485\linewidth]{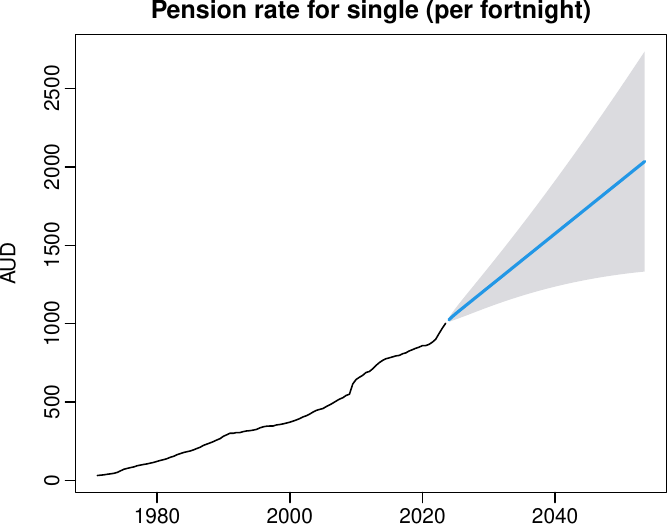}
    \quad 
    \includegraphics[width=0.485\linewidth]{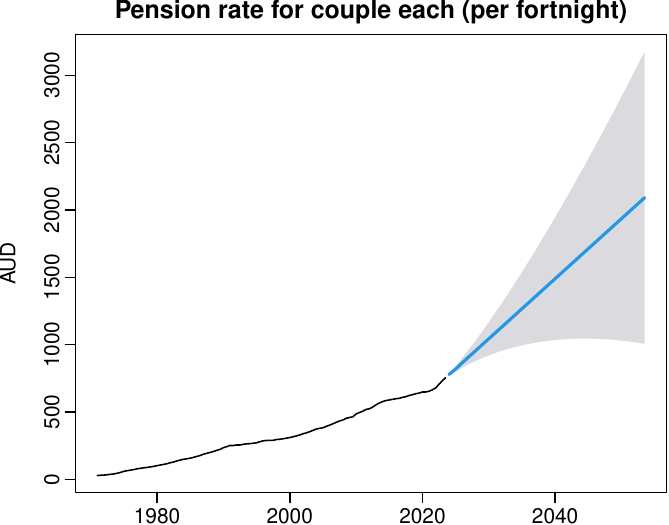}
    \caption{\small{Maximum basic age pension rates for singles and couples (black line) in Australia between 1971 and 2023. The forecasts of age pension rates until 2051 are represented by blue curves, with the corresponding 95\% pointwise prediction intervals indicated by the shaded area.}}
    \label{fig:12}
\end{figure}

The Australian Government has consistently increased age pension rates since the 1970s to align with inflation and rising living costs. We obtained the maximum basic age pension rates for singles and couples between 1971 and 2023 from the \cite{socialsecurity}. Two rate adjustments occur annually during the considered time, leading to both time series of payout rates exhibiting clear upward trends, as illustrated in Figure~\ref{fig:12}. We apply the univariate time series forecasting method to pension rates via the automatic algorithm of \cite{hyndman2008automatic}. The forecasts in Figure~\ref{fig:12} inherit the increasing trends of age pension rates over recent decades. This study does not consider other minor age pension supplements (e.g., Energy Supplement provided by the Department of Social Services) because of their low payment value and the restricted number of covered recipients. 

Combining forecasts of life expectancies and the maximum basic pension rates can provide insight into the distribution of age pension welfare in Australia. We consider single and couple residents in each state and territory across Australia who are eligible for the age pension starting from the following years: 2022, 2027, 2032, 2037, 2042, 2047, and 2051. For ease of comparison, Table~\ref{tab:1} shows the present values in 2023 of the projected future age pension payments discounted using the average real interest rates between 1971 and 2023 obtained from \cite{worldbank}. 
\begin{center}
\tabcolsep 0.166in
\begin{longtable}{@{}lrrrrrrr@{}}
\caption{\small{The projected lifetime age pension income (present value in 2023 \$AUD) for single and couple residents in each state and territory across Australia.}}\label{tab:1} \\
\toprule  
State & 2022 & 2027 & 2032 & 2037 & 2042 & 2047 & 2051 \\ \midrule
\endfirsthead
\toprule
State & 2022 & 2027 & 2032 & 2037 & 2042 & 2047 & 2051 \\ \midrule
\endhead
\hline \multicolumn{8}{r}{{Continued on next page}} \\
\endfoot
\endlastfoot
& \multicolumn{7}{c}{\textbf{Female}} \\
	ACT 	 & \$438,493 & \$495,795 & \$549,133 & \$626,843 & \$677,411 & \$756,061 & \$819,707 \\ 
  	NSW & \$462,209 & \$521,487 & \$577,497 & \$657,052 & \$710,340 & \$791,272 & \$856,699 \\ 
 	QLD  & \$457,305 & \$515,833 & \$570,724 & \$649,725 & \$702,248 & \$782,433 & \$847,510 \\ 
  	WA    & \$474,832 & \$535,064 & \$591,366 & \$671,427 & \$725,455 & \$806,728 & \$872,754 \\ 
  	NT     & \$320,211 & \$382,115 & \$443,118 & \$533,255 & \$596,749 & \$692,299 & \$770,663 \\ 
  	VIC    & \$469,705 & \$530,167 & \$586,832 & \$667,340 & \$721,588 & \$803,165 & \$869,283 \\ 
  	TAS   & \$409,918 & \$459,645 & \$506,256 & \$578,013 & \$621,630 & \$694,680 & \$754,036 \\ 
  	SA     & \$455,325 & \$512,480 & \$565,655 & \$643,140 & \$693,958 & \$772,779 & \$836,832 \\ 
\midrule
& \multicolumn{7}{c}{\textbf{Male}} \\
  	ACT & \$375,188 & \$428,383 & \$478,460 & \$554,904 & \$603,183 & \$681,580 & \$745,455 \\ 
  	NSW & \$384,412 & \$439,531 & \$492,417 & \$571,527 & \$622,977 & \$704,277 & \$770,361 \\ 
  	QLD & \$377,661 & \$432,526 & \$484,654 & \$563,112 & \$613,712 & \$694,224 & \$759,696 \\ 
  	WA & \$392,687 & \$449,408 & \$503,310 & \$583,165 & \$635,553 & \$717,545 & \$784,314 \\ 
  	NT & \$256,324 & \$301,194 & \$349,002 & \$426,527 & \$475,635 & \$559,267 & \$628,433 \\ 
  	VIC & \$396,663 & \$454,587 & \$508,907 & \$589,041 & \$641,780 & \$723,747 & \$790,338 \\ 
  	TAS & \$331,089 & \$378,685 & \$422,762 & \$494,194 & \$536,425 & \$610,315 & \$670,738 \\ 
  	SA & \$367,231 & \$421,124 & \$469,890 & \$545,127 & \$592,188 & \$669,827 & \$733,084 \\
\midrule
& \multicolumn{7}{c}{\textbf{Couple Each}} \\
  	ACT & \$314,305 & \$382,393 & \$448,921 & \$542,376 & \$607,318 & \$704,223 & \$783,584 \\ 
  	NSW & \$322,901 & \$393,208 & \$462,896 & \$559,519 & \$628,126 & \$728,553 & \$810,658 \\ 
  	QLD & \$316,611 & \$386,413 & \$455,123 & \$550,835 & \$618,370 & \$717,768 & \$799,065 \\ 
  	WA & \$330,633 & \$402,805 & \$473,809 & \$571,526 & \$641,369 & \$742,810 & \$825,842 \\ 
  	NT & \$206,780 & \$261,746 & \$321,528 & \$411,678 & \$474,493 & \$574,005 & \$657,171 \\ 
  	VIC & \$334,371 & \$407,864 & \$479,444 & \$577,609 & \$647,926 & \$749,474 & \$832,408 \\ 
  	TAS & \$273,690 & \$334,650 & \$393,626 & \$480,214 & \$537,528 & \$628,160 & \$702,729 \\ 
  	SA & \$306,906 & \$375,368 & \$440,377 & \$532,333 & \$595,783 & \$691,632 & \$770,173 \\ 
\bottomrule
\end{longtable}
\end{center}

From Table~\ref{tab:1}, NT and TAS residents are expected to receive much lower age pension income throughout their retirement than the remaining six states and territories. The below-average age pension welfare distributed to retirees in NT and TAS underscores the substantial income disparity in these areas, as indicated by the lowest Gini coefficients of NT and TAS among the eight states and territories in Australia \citep{kennedy2017does}. Most Australians nowadays partially or fully rely on the age pension as the primary source of income in retirement. In 2023, around 92\% of Australians aged 65 and over who received income support were beneficiaries of the age pension \citep{AIHW23}. The inequality in the age-pension welfare distribution over life long may put the extra cost of living pressures on older Australians in regional areas since the negative effects of increasing pension ages are found to disproportionately affect poorer households \citep{morris2022unequal}. Hence, we recommend that the federal government and governments in regional and rural areas consider providing retirees with low socioeconomic status with extra financial and health support.

Lifetime age pension projections were completed on an annual basis for the years between 2023 and 2032. The results also indicate that NT and TAS residents will likely receive much lower lifetime age pension income than pensioners in NSW and VIC. Due to space constraints, the annual age-pension projection results are available at \url{https://yy135.shinyapps.io/shiny/}.

\section{Conclusion} \label{sec:7}

Australia faces a pressing issue shared by many other nations -- an aging population. This aging population is due to people living longer than ever before, combined with the retirement of the baby boomer generation. As a result, the cost of providing age pensions has increased substantially, and this trend is likely to continue unless significant changes are made to current pension schemes. To address this issue, the Australian Government has proposed a gradual increase in the pension age from 65 to 67 by July 2023, possibly increasing to 70 by 2035. However, our analysis suggests that such a rapid increase in pension age may not be sustainable unless the goal is to reduce the OADR to below 2018 levels.

Our paper proposes a statistical approach to model and forecast sub-national populations by age and sex. Population forecasts at a more granular sub-national level are needed to facilitate local-level decisions, such as health and regional migration policies. These forecasts can also reveal patterns of vulnerable groups and track the effects of policy responses. By aggregating the sub-national population forecasts to the national level, we measure the effects of changes in the pension age on the OADR, which we take as a crude measure of the financial burden of the age pension. We have applied stochastic models to forecast the future population age structure and then used those future age structure projections to determine the optimal pension age scheme that will ensure a consistent and desirable level of OADR.

We have employed a reverse methodology to establish a pension age plan to achieve a targeted OADR value. Our approach assumes that the optimal OADR level is approximately 23\%, recorded in 2018. Our analysis, which involved historical population data from 1921 to 2021 and making population forecasts, has revealed that the total population is expected to increase significantly, and the aging trend will continue. Based on our findings, we recommend that the pension age should not be fixed but should instead increase at a rate that maintains a stable OADR level feasible for working Australians. 

Through the research conducted in this paper, it has been demonstrated that the decision to raise the pension age to 67 has had a successful impact on halting the upward trend in OADR values. While this may have provided a short-term solution, more is needed to ensure a stable OADR in the long run, especially considering the continuous changes in the population's age profile. Therefore, it is suggested that the pension age be regularly reviewed using the proposed statistical approach to maintain a sustainable OADR. 

Our research has shown that the projected pension age is expected to increase steadily over the next few years. Specifically, it is estimated to rise to 67 by 2024, 68 by 2029, 69 by 2037, and 70 by 2043. Despite this expected rise, our analysis indicates that the OADR will remain stable at approximately 23\%. However, one must also consider economic factors, such as the impact of the superannuation guarantee and other measures to relieve the financial burden of the age pension. Moreover, not all people over the pension age threshold are eligible to receive it, and the number of recipients may decrease as more citizens gain access to superannuation. In addition, many self-funded retirees withdraw from paid employment before reaching the pension age, which could impact labor force participation. Therefore, raising the pension age may not impact the OADR as much as previously estimated. Finally, it is important to consider the effect of changes in per capita GDP, as a decline in this measure may increase the fiscal burden of the pension, even if the OADR remains stable. For reproducibility, the computer \Rlogo \ code is available at \url{https://github.com/csz16/Hamilton-Perry-method}.

Although the dataset timeframe is until 2021, including two years of data after the pandemic, we have yet to consider the impact of COVID-19 in this analysis since it is too early to estimate its effect on future age profiles accurately. We made no assumptions about mortality using the HP model. It is known that COVID-19 has mainly affected older people's mortality rates, which should decrease the OADR. However, Australia's limited number of deaths caused by COVID-19 suggests that the pandemic's effect is unlikely to alter our recommendations significantly.

In our calculation, we acknowledge that many factors should be considered, and they do not factor into living expenses in different states. In addition, we do not consider the type of job, as such data are not publicly available in Australia. Our projection and findings consider a demographic factor based on the average remaining life expectancy. Living longer does not imply a healthy life; as a result, it is interesting to study the forecast cost of aging care to the government \citep[see, e.g.,][]{CSL+24}.

\section*{Acknowledgment}

The authors would like to acknowledge the insightful comments from two reviewers, which led to a much-improved paper, and are grateful for Dr. Tom Wilson's comments. This research is funded by an internal grant from the Data Horizons Consilience Research Center at Macquarie University.

\bibliographystyle{agsm}
\bibliography{HP_pension.bib}

\end{document}